\documentstyle[12pt,epsf]{article}

\begin{document}

\pagestyle{empty}

\begin{flushright}
%%%{\bf TSU-HEPI-2000-05}\\
{\bf MZ--TH/02--10}
\end{flushright}

\vglue 2cm

\begin{center} \begin{Large} \begin{bf}
One-loop corrections to four-point functions with two external massive
fermions and two external massless partons.
\end{bf} \end{Large} \end{center}
\vglue 0.35cm
{\begin{center}
J.G.\ K\"{o}rner$^{a,\dag}$ and
Z.\ Merebashvili$^{b,\ddag}$
\end{center}}
\parbox{6.4in}{\leftskip=1.0pc
{\it a.\ Institut f\"{u}r Physik, Johannes Gutenberg-Universit\"{a}t,
         D-55099 Mainz, Germany}\\
\vglue -0.25cm
{\it b.\ High Energy Physics Institute,
Tbilisi State University, 380086 Tbilisi, Georgia}
}
\begin{center}
\vglue 1.0cm
\begin{bf} ABSTRACT \end{bf}
\end{center}
\vglue 1.0cm
{\rightskip=1.5pc
\leftskip=1.5pc
\tenrm\baselineskip=12pt
 \noindent
We present a complete set of one-loop matrix elements relevant for
the hadroproduction of heavy quarks in next-to-leading order employing
dimensional regularization to isolate ultraviolet and soft
divergences. All results of the perturbative calculation
are given in detail. These one-loop matrix elements 
can also be used  as input in the determination of the corresponding
next-to-leading order cross sections for heavy flavor photoproduction
and in photon-photon reactions, as well as for any of the relevant crossed
processes. Our results are tested against the results of other related 
studies in which unpolarized and longitudinally polarized processes 
were considered.
\vglue 2cm
PACS number(s): 12.38.Bx, 13.85.-t, 13.85.Fb, 13.88.+e
}

\renewcommand{\thefootnote}{\dag}
\footnotetext{e-mail address: koerner@thep.physik.uni-mainz.de}
\renewcommand{\thefootnote}{\ddag}
\footnotetext{e-mail address: mereb@sun20.hepi.edu.ge}

\newpage

\pagestyle{plain}
\setcounter{page}{1}

%\vglue .3cm
\renewcommand{\theequation}{1.\arabic{equation}}

\begin{center}\begin{large}\begin{bf}
I. INTRODUCTION
\end{bf}\end{large}\end{center}
\vglue .3cm

The production of heavy quarks is important for our understanding of 
nature.
Intensive experimental studies of a heavy quark production in
various reactions involving unpolarized initial particles are presently
being carried out. However, not less important are analogous studies when
initial particles are polarized, either longitudinally or transversely.
Experiments on longitudinally polarized initial particles are taking
place $[$\ref{compass}$]$, are being planned $[$\ref{slac}$]$ and proposed
$[$\ref{desy}$]$. These will deal with many aspects of polarized
reactions, e.g. the problem of the polarized gluon structure function 
$g_2$,
the validity of the Drell-Hearn-Gerasimov (DHG) sum rule in QCD, etc...

At the leading order (LO) Born term level, various heavy quark production 
mechanisms have been studied some time ago. 
However, the importance of knowing the next-to-leading order (NLO)
corrections cannot be overemphasized.
Unpolarized NLO corrections for heavy quark
hadroproduction were first presented in $[$\ref{Nason},\ref{Been}$]$, and 
in $[$\ref{Ellis},\ref{Smith}$]$ 
for photoproduction. Corresponding polarized results
were calculated in $[$\ref{Bojak1}$]$ and 
$[$\ref{Bojaka},\ref{Bojakb},\ref{MCGa},\ref{MCGb}$]$.
In all these papers cross sections were obtained by folding in the Born 
term matrix
elements from the very beginning. Analytical results for the so called
``virtual plus soft'' terms were presented in
$[$\ref{Been},\ref{Smith},\ref{Bojakb}$]$ for the photoproduction and 
unpolarized hadroproduction of heavy quarks. Complete analytic results
for the polarized and unpolarized photoproduction, including real
bremsstrahlung, can be found in $[$\ref{MCGb}$]$.

Let us emphasize the importance of knowing one-loop matrix elements, which
contain the full spin information of the relevant subprocess. 
When the one-loop contributions are folded with the Born term 
contributions and spin summed, as in a NLO rate calculation, the 
information 
on the spin content of the one-loop contribution is lost and cannot be 
reconstructed from the rate expressions. On the other hand, having 
expressions for matrix elements allows one 
to easily derive the one-loop contributions to partonic cross section 
including any polarization of the 
incoming or outgoing particles. Also, it allows one to obtain any of the 
crossed
processes, including the ones with a heavy incoming particle, that are
needed for the different versions of variable flavor number 
schemes, for the 
direct checking of the DHG sum rule on the partonic level and
possibly for doing future parametrizations for structure functions. And
all this without doing calculations from the beginning.
This work presents detailed results on a NLO calculation of partonic
matrix elements for the set of one-loop Feynman graphs present in 
hadroproduction
of heavy flavors, separately for every Feynman diagram in order to
facilitate the use of the results for other relevant processes that
differ by color factors.

The one-loop results presented in this paper are an important input for 
part of the next-to-next-to-leading (NNLO) order calculation of the 
perturbative 
corrections to heavy flavor production. Last but not least, the imaginary 
parts of the one-loop amplitudes 
presented here are a necessary ingredient for the calculation of 
$T$-odd observables which are known to be fed by the NLO absorptive (or
imaginary) parts of the one-loop contributions to a given process.

The subprocesses that receive one-loop corrections which are 
considered in this paper proceed through the following two partonic
channels:
\begin{equation}
\label{gluglu}
g + g \rightarrow Q + \overline Q,
\end{equation}
where $g$ denotes a gluon and $Q (\overline {Q})$ denotes a heavy quark 
(antiquark), and
\begin{equation}
\label{qbarq}
q + \bar{q} \rightarrow Q + \overline Q,
\end{equation}
where $q (\bar{q})$ is a light massless quark (antiquark).

We note that the Abelian part of the NLO result for (\ref{gluglu}) 
provides the NLO corrections to heavy flavor production by two on-shell 
photon collisions
\begin{equation}
\label{gamgam}
\gamma + \gamma \rightarrow Q + \overline Q,
\end{equation}
with the appropriate color factor substitutions. 
The results for (\ref{gluglu}) can be also used to determine corresponding 
amplitudes for heavy flavor photoproduction
\begin{equation}
\label{gamglu}
\gamma + g \rightarrow Q + \overline Q.
\end{equation}
We mention that the partonic processes (\ref{gluglu}) and (\ref{qbarq}) 
are needed for the 
calculation of the contributions of single- and double-resolved photons in 
the photonic processes (\ref{gamgam}) and (\ref{gamglu}).

Cross sections for the process (\ref{gamgam})
have already been determined in $[$\ref{Mirkes},\ref{Drees},\ref{KMC}$]$ 
for unpolarized and in $[$\ref{KMC},\ref{JT}$]$ for polarized
initial photons. Note that the authors of $[$\ref{JT}$]$ used a 
nondimensional regularization scheme to regularize the poles of divergent 
integrals. In the papers $[$\ref{Mirkes},\ref{JT}$]$ analytic 
results were presented for ``virtual plus soft'' contributions alone.
We also note that complete analytic results including hard gluon 
contributions can be found only in $[$\ref{KMC}$]$.
The reaction (\ref{gamgam}) will be investigated at future linear
colliders. NLO corrections for the heavy quark production cross section
(\ref{gamgam}) are of interest in themselves as they represent an
irreducible background to the intermediate Higgs boson searches for
Higgs masses in the range of 90 to 160 GeV.

The paper is organized as follows. 
Section~II contains an outline of our general approach as well as matrix 
elements of the gluon fusion subprocess for the self-energy and vertex
contributions including their renormalization. 
In Section~III we discuss the one-loop contributions to the four box 
diagrams in the same gluon-gluon subprocess and give a detailed 
description of our global checks on gauge invariance for our results. 
Section~IV presents analytic results on the quark-antiquark subprocess 
(\ref{qbarq}). 
Whereas Sections~II--IV deal with the real (or dispersive) parts of the 
one-loop amplitudes we turn to discuss their imaginary (or absorptive) 
parts in Section~V. Section~V also includes a discussion on how one may 
obtain the corresponding absorptive parts in different crossed channels.
Our main results are summarized in Section~VI. 
Finally, in two Appendices we present results for various coefficient 
functions completing our determination of NLO matrix elements.

%\newpage
\renewcommand{\theequation}{2.\arabic{equation}}
\setcounter{equation}{0}
\vglue 1cm
\begin{center}\begin{large}\begin{bf}
II. CONTRIBUTIONS OF THE TWO- AND THREE-POINT 
FUNCTIONS TO GLUON FUSION
\end{bf}\end{large}\end{center}
\vglue .3cm

The Born and the one-loop contributions to the gluon fusion partonic 
reaction $g(p_1)+g(p_2)\rightarrow Q(p_3) + \overline {Q}(p_4)$ are shown 
in Figs.~1--3.
In this section we discuss our calculation of the self-energy and vertex 
graphs that contribute to the above subprocess.
With the 4-momenta $p_i (i=1,...,4)$ as indicated in the Fig.~1 and with 
$m$ the heavy quark mass we define:
\begin{equation}
s\equiv (p_1+p_2)^2, {\rm \hspace{.3in}}  t\equiv T-m^2
\equiv (p_1-p_3)^2-m^2,
{\rm \hspace{.3in}}  u\equiv U-m^2\equiv (p_2-p_3)^2-m^2.
\end{equation}

To isolate ultraviolet (UV) and infrared/collinear (IR/M) divergences
we have carried out all our calculations in both conventional 
regularization schemes, namely the standard dimensional regularization 
scheme (DREG) $[$\ref{DREG}$]$ 
and the dimensional reduction scheme (DRED) $[$\ref{DRED}$]$. In what 
follows, we present results for the DREG, as well as the difference 
$\Delta$=DRED-DREG.
A brief characterization of the two regularization schemes is the 
following: In DREG both tensorial structures (e.g. gamma matrices, metric 
tensors, etc...) and momenta 
are continued to $n\neq 4$, while in DRED only momenta are 
continued to $n\neq 4$ whereas the tensorial structures are those of 
$n=4$.

First of all we note that in general the matrix elements for all 
the Feynman diagrams in the gluon fusion subprocess are written in the form
\begin{equation}
M = \epsilon_{\mu}(p_1) \epsilon_{\nu}(p_2) \bar{u}(p_3) M^{\mu\nu} v(p_4),
\end{equation}
However, for the purposes of brevity, we will present our results in terms 
of the amplitudes $M^{\mu\nu}$ omitting the polarization vectors and Dirac 
spinors.
Of course, their presence is implicitly understood throughout this paper 
in that the mass shell conditions 
$p_1^{\mu}\epsilon_{\mu}(p_1)=0$ and $\not {\rm \hspace{-.03in}} 
p_3 u(p_3)=m u(p_3)$ etc. are 
being used to simplify $M^{\mu\nu}$       $^1$. 
Furthermore, $M^{\mu\nu}$ for all the 
one-loop graphs considered in this paper contains a common factor due to 
the one-loop integration. Define the quantity
\renewcommand{\thefootnote}{\arabic{footnote}}
\addtocounter{footnote}{1} 
\footnotetext{According to the discussion in $[$\ref{Slaven}$]$ this 
implies that, when further processing our LO and one-loop results in 
cross section calculations by 
folding in the appropriate amplitudes, one may use the Feynman gauge for 
the spin sums of polarization vectors. At the same time ghost
contributions associated with external gluons have to be omitted.} 
\begin{equation}
\label{ceps}
C_{\varepsilon}(m^2)\equiv\frac{\Gamma(1+\varepsilon)}{(4\pi)^2}
\left(\frac{4\pi\mu^2}{m^2}\right)^\varepsilon .
\end{equation}
We will omit from all of our one-loop $M^{\mu\nu}$ amplitudes the 
common factor
\begin{equation}
\label{common}
{\rm \cal C} = g^4 C_{\varepsilon}(m^2),
\end{equation}
where $g$ is the renormalized coupling constant.

For our analysis of matrix elements it is important to describe various 
crossed heavy flavor production channels. 
We make it clear from the outset that an additional u-channel set of 
graphs, that topologically differ from the t-channel ones, are obtained by 
interchange of bosonic lines (not momenta). 
In particular, for calculational purposes, we will always be
relating t- and u-channel Feynman diagrams by the following procedure:
\begin{equation}
\label{tu}
{\cal M}_t \leftrightarrow {\cal M}_u   \equiv
\{     a\leftrightarrow b,  {\rm \hspace{.2in}}
p_1\leftrightarrow p_2,  {\rm \hspace{.2in}}  \mu\leftrightarrow \nu  \},
\end{equation}
with $a,b$ color indices of bosons and
where all three interchanges are performed {\it simultaneously}. Note that 
the second interchange in (\ref{tu}) implies also the interchange 
$t\leftrightarrow u$, but not vice versa. One case, involving two 
vertex diagrams, when the above transformation (\ref{tu}) does not 
correspond to ``true'' u-channel topologies, is discussed below. 
In general, when speaking about t-u symmetry of given amplitudes, we 
will imply invariance of those amplitudes under the transformations 
(\ref{tu}).

We start by writing down matrix elements for the leading order Born terms. 
For the t-channel gluon fusion subprocess (first graph in Fig.~1) we have:
\[
B_t^{\mu\nu} = -i T^b T^a \gamma^{\mu}(   {\rm \hspace{-.1in}}
\not p_3 - {\rm \hspace{-.1in}}  \not p_1 + m) \gamma^{\nu}   /t,
\]
where $T^b$ and $T^a$ are generators ($T^a=\lambda^a/2$, 
$a=1,...,8$ and the $\lambda^a$ are Gell-Mann matrices) that define the 
fundamental representation of the Lie algebra of the color SU(3) group.
Analogously, for the u- and s-channels we have, respectively,
\begin{eqnarray}
\nonumber
B_u^{\mu\nu} &=& -i T^a T^b \gamma^{\nu}( {\rm \hspace{-.1in}}  \not p_3 
              - {\rm \hspace{-.1in}}  \not p_2 + m)   \gamma^{\mu} /u,   \\
\nonumber
B_s^{\mu\nu} &=& i(T^a T^b - T^b T^a) C_3^{\mu\nu\sigma} \gamma_{\sigma}/s,
\end{eqnarray}
where the tensor $C_3^{\mu\nu\sigma}$ is defined according to the Feynman 
rules for the three-gluon coupling. We have omitted a common factor 
$g^2$ in the Born amplitudes. 
Acting with Dirac spinors on the above Born matrix elements from the left 
and the right and using the effective relations $p_1^{\mu}=p_2^{\nu}=0$, 
as remarked on before, we arrive at the following expressions 
for the leading order matrix elements:
\begin{eqnarray}
\nonumber
B_t^{\mu\nu} &=& i T^b T^a ( \gamma^{\mu} {\rm \hspace{-.1in}} \not p_1 
              \gamma^{\nu} - 2 p_3^{\mu} \gamma^{\nu} )/t;          \\
B_u^{\mu\nu} &=& i T^a T^b ( 2 p_4^{\mu} \gamma^{\nu} - \gamma^{\nu} 
{\rm \hspace{-.1in}}                  \not p_1 \gamma^{\mu} )/u
= i T^a T^b ( \gamma^{\nu} {\rm \hspace{-.1in}} \not p_2
              \gamma^{\mu} - 2 p_3^{\nu} \gamma^{\mu} )/u;    \\
\nonumber
B_s^{\mu\nu} &=& 2i(T^a T^b - T^b T^a)(g^{\mu\nu} {\rm \hspace{-.1in}} 
           \not p_1 + p_2^{\mu} \gamma^{\nu} - p_1^{\nu} \gamma^{\mu})/s.
\end{eqnarray}

Next we proceed with the description of the two-point contributions to the 
matrix element of the subprocess (\ref{gluglu}). 
But before we turn to the two-point functions one should mention that our 
choice of renormalization scheme will be a {\it fixed flavor} scheme 
throughout this paper. This implies that we have a total number of flavors 
$n_f=n_{lf}+1$, where $n_{lf}$ is the number of light (e.g. massless) 
flavors plus one produced heavy flavor, with only $n_{lf}$ light 
flavors involved/active in the $\beta$ function for the running a QCD 
coupling $\alpha_s$ and in the splitting functions that determine the 
evolution of the structure functions.
When having massless particles in the loops we are using the standard 
$\overline {\rm MS}$ scheme, while the contribution of a heavy quark loop 
in the gluon self-energy with on-shell external legs is subtracted out 
entirely.

Consider first the two t-channel self-energy graphs (2d2) and (2d3) 
with external legs on-shell (note that 
in the graph numeration the first number identifies the figure which 
the given diagram refers to). 
These graphs are very important as they determine the renormalization 
parameters in the quark sector. 
Throughout this paper we use the so called on-shell prescription for the 
renormalization of heavy quarks, the essential ingredients of which we 
describe in the following. 
When dealing with massive quarks one has to choose a parameter to 
which one renormalizes the heavy quark mass. 
It is natural to choose a quark pole mass for such a parameter - the only 
``stable'' mass parameter in QCD. The condition on the renormalized 
heavy quark self-energy $\Sigma_{r}(\not p)$ is 
\begin{equation}
\label{massren}
\Sigma_{r}(\not p)|_{\not p=m} = 0,
\end{equation}
which removes the singular internal propagator in these self-energy 
diagrams. The above condition determines the mass renormalization constant 
$Z_m$. For the wave function renormalization we have used the usual 
condition (see e.g. Ref.~$[$\ref{Ellis}$]$)
\begin{equation}
\label{waveren} 
\frac{\partial}{\partial{\rm \hspace{-.07in}}\not p} 
\Sigma_{r}(\not p)|_{\not p=m} = 0,                   
\end{equation}
which fully determines the wave function renormalization constant $Z_2$. 
Since the condition (\ref{waveren}) is not mandatory in general, there is a 
freedom in determining the constant $Z_2$. Therefore, we will list 
our expressions for these constants. 
In DREG we arrive at
\begin{equation}
Z_m = 1 - 3 g^2 C_F C_{\varepsilon}(m^2) \left( \frac{1}{\varepsilon'} + 
\frac{4}{3} \right),
{\rm \hspace{.4in}}
Z_2 = 1 - g^2 C_F C_{\varepsilon}(m^2) \left( \frac{1}{\varepsilon'} + 4 +
\frac{2}{\varepsilon} \right).
\end{equation}
And in DRED we obtain 
\begin{equation}
Z_m = 1 - 3 g^2 C_F C_{\varepsilon}(m^2) \left( \frac{1}{\varepsilon'} +
\frac{5}{3} \right),
{\rm \hspace{.4in}}
Z_2 = 1 - g^2 C_F C_{\varepsilon}(m^2) \left( \frac{1}{\varepsilon'} + 5 +
\frac{2}{\varepsilon} \right),
\end{equation}
where $C_F$=4/3 and where we use $1/\varepsilon'$ 
to indicate which terms are of ultraviolet origin.
Now we are in a position to write our results for the (2d2) and (2d3) quark 
self-energy diagrams. In DREG we have:
\begin{equation}
M_{\rm (2d2)}^{\mu\nu} = M_{\rm (2d3)}^{\mu\nu} = - C_F B_t^{\mu\nu} 
           \left( \frac{1}{\varepsilon'} + 4 \ln\frac{m^2-p_3^2}{m^2} 
\right),
\end{equation}
where we clearly see the soft divergence in the logarithm that diverges 
on the mass-shell. Apparently, this very same logarithm appears also when 
calculating the wave function renormalization constant $Z_2$. This means 
that on-shell mass renormalization does not actually completely remove the 
divergent propagator, but rather transforms the strong linear divergence to 
the ``softer'' logarithmic divergence, which can be dealt with in QCD. 
Indeed, one can rigoriously prove that there exists the following effective 
correspondence relation:
\begin{equation}
\ln\frac{m^2-p^2}{m^2} \mid_{p^2=m^2} \rightarrow 
\frac{1}{2\varepsilon} + \frac{1}{1-2\varepsilon} = \frac{1}{2\varepsilon} 
+ 1 + {\rm \cal O}(\varepsilon).
\end{equation}
With this in mind we obtain final results for the two self-energy 
graphs {\it after mass renormalization was carried out}:
\begin{equation}
\label{grd}
{\rm \hspace{-1.4in}   {DREG:}    \hspace{.8in}}
M_{\rm (2d2)}^{\mu\nu} = M_{\rm (2d3)}^{\mu\nu} = - C_F B_t^{\mu\nu}
           \left( \frac{1}{\varepsilon'} + 4 + \frac{2}{\varepsilon}
\right),
\end{equation}
\begin{equation}
\label{grddred}
{\rm \hspace{-1.4in}   {DRED:}   \hspace{.8in}}
M_{\rm (2d2)}^{\mu\nu} = M_{\rm (2d3)}^{\mu\nu} = - C_F B_t^{\mu\nu}
           \left( \frac{1}{\varepsilon'} + 5 + \frac{2}{\varepsilon}
\right),
\end{equation}
and the difference between the two regularization schemes is
\begin{equation}
\Delta(2d2) = \Delta(2d3) = - C_F B_t^{\mu\nu}.
\end{equation}
One notices that the effect of the wave function renormalization consists 
of a complete removal of the quark self-energy diagrams with external legs 
on-shell, as it is required by the second condition (\ref{waveren}).
We can also write the contribution of the quark self-energy with external 
legs off-shell, graph (2d1), after addition of the mass renormalization 
counterterm:
\begin{eqnarray}
\label{grd1}
M_{\rm (2d1)}^{\mu\nu} &=& C_F B_t^{\mu\nu}
\left( -1/\varepsilon' - t/T + \ln(-t/m^2) (4 t/T + t^2/T^2 - 4)
\right)      \\
\nonumber
&-& i C_F T^b T^a m \gamma^{\mu}\gamma^{\nu} \left( 1-2 \ln(-t/m^2) 
- \ln(-t/m^2) t/T
\right) /T.
\end{eqnarray}
The difference between the DRED and DREG results is
\begin{equation}
\Delta(2d1) = - C_F B_t^{\mu\nu}.
\end{equation}

The remaining quark self-energy diagrams (3i1) and (3i2) with external 
on-shell legs are derived in analogy to the ones considered above:
\begin{equation}
\label{gri}
M_{\rm (3i1)}^{\mu\nu} = M_{\rm (3i2)}^{\mu\nu} = - C_F B_s^{\mu\nu}
           \left( \frac{1}{\varepsilon'} + 4 + \frac{2}{\varepsilon}
\right),
\end{equation}
\begin{equation}
\Delta(3i1) = \Delta(3i2) = - C_F B_s^{\mu\nu}.
\end{equation}

Concerning the gluon self-energy graphs (2e1) and (2e2) with external legs 
on-shell, the only nonvanishing 
contribution they receive from the loop with internal heavy quarks 
is given by 
\begin{equation}
\label{gre}
M_{\rm (2e1)}^{\mu\nu} = M_{\rm (2e2)}^{\mu\nu} = - B_t^{\mu\nu} 
\frac{1}{\varepsilon'} \,\, \frac{2}{3}.
\end{equation}
However, these contributions are explicitly subtracted (together with the 
logarithmic term $\ln(\mu^2/m^2)$ coming from the common factor 
$C_{\varepsilon}(m^2)$, see eqs.~(\ref{ceps}) and (\ref{common})) in the 
on-shell renormalization 
prescription, in order to avoid the appearance of the large mass logarithms 
from the gluon self-energy diagrams with off-shell external legs in the low 
energy limit. 
Therefore, due to the UV counterterm that subtracts that very same loop 
with heavy quarks, there are no finite contributions to the matrix element 
from these diagrams. 
However, at the same time this counterterm 
introduces the pole terms from the light quark loop sector that are 
needed to cancel soft and collinear poles from the other 
parts of the amplitude, e.g. from the real bremsstrahlung part. This 
indicates that in practice it is very hard to completely disentangle UV 
and IR/M poles in heavy flavor production and in most cases one obtains a 
mixture of both instead. 

For the reasons specified above it is convenient to present 
a gauge field renormalization constant $Z_3$, used for the gluon 
self-energy subtraction:
\begin{eqnarray}
\nonumber
Z_3 &=& 1 + \frac{g^2}{\varepsilon'} \left\{ (\frac{5}{3} N_C - \frac{2}{3} 
n_{lf}) C_{\varepsilon}(\mu^2) - \frac{2}{3} C_{\varepsilon}(m^2)\right\}  
\\
&=& 1 + \frac{g^2}{\varepsilon'}\left\{ (b - 2N_C) C_{\varepsilon}(\mu^2) - 
\frac{2}{3} C_{\varepsilon}(m^2)\right\},
\end{eqnarray}
with the QCD beta function $b=(11 N_C - 2 n_{lf})/6$ containing only light 
quarks. $N_C=3$ is the number of colors.

Similarly to the diagrams (2e1) and (2e2), diagrams (3j1) and (3j2) also 
vanish 
due to the explicit decoupling of the heavy quarks in our subtraction 
prescription. However, instead of doing renormalizations 
separately for each Feynman diagram, one can chose to employ the 
renormalization group invariance of the cross section and do only 
a mass and coupling constant renormalization. In that case, 
knowing the results for gluon self-energies turns out to be useful in 
checking the complete cancellation of UV poles by just rescaling the 
coupling constant in the LO terms $g_{\rm bare}\rightarrow Z_g g$:
\begin{equation}
\label{grj}
M_{\rm (3j1)}^{\mu\nu} = M_{\rm (3j2)}^{\mu\nu} = - B_s^{\mu\nu}
\frac{1}{\varepsilon'} \,\, \frac{2}{3}.
\end{equation}

Finally we arrive at the gluon-self energy graph (3h), which 
contains the off-shell gluon self-energy loop that is used for the 
derivation of the 
renormalization constant $Z_3$. We have evaluated the internal loop in the 
Feynman gauge. Since it is explicitly gauge invariant, we 
should arrive at the same result in any other gauge. In our result we 
show separately the gauge invariant pieces for gluon plus ghost, 
light quarks and a heavy quark flow inside the loop:
\begin{equation}
\label{grh}
M_{\rm (3h)}^{\mu\nu} = B_s^{\mu\nu} \left\{ \left[ N_C \left( 
\frac{1}{\varepsilon'} \,\, \frac{5}{3} + \frac{31}{9} \right) - n_{lf} 
\left( \frac{1}{\varepsilon'} \,\, \frac{2}{3} + \frac{10}{9} \right) 
\right] 
\frac{C_{\varepsilon}(-s)}{C_{\varepsilon}(m^2)}
- \frac{1}{\varepsilon'} \,\, \frac{2}{3} \,\, I 
\right\},
\end{equation}
with
\begin{equation}
\label{integ}
I = 1 + \varepsilon' \, \left\{ \frac{5}{3} + \frac{4m^2}{s} + 
\left( 1+\frac{2m^2}{s} \right) \beta \ln(x) \right\}.
\end{equation}
In (\ref{integ}) we have made use of the definitions 
\begin{equation}
\label{varx}
\beta \equiv \sqrt{1-4m^2/s},    {\rm \hspace{.6in}}
x \equiv \frac{1-\beta}{1+\beta},
\end{equation}
where $\beta$ is the velocity of the heavy quark.

We emphasize that in the last term of (\ref{grh}) the $Z_3$ counterterm 
together with the UV pole will remove also the $\ln(\mu^2/m^2)$ 
contribution, 
while $\ln(-\mu^2/s)$ from the first two terms in (\ref{grh}) will be left 
unsubtracted. Expression (\ref{grh}) is obtained for the 
physical condition $s\ge 4m^2$ for producing a heavy quark pair. 
However, if one is interested in the case when $s < 4m^2$, then in 
(\ref{integ}) one should make the replacement
\begin{equation}
\beta \ln(x) \rightarrow -2 \sqrt{4m^2/s-1} \,\,
\arctan{\frac{1}{\sqrt{4m^2/s-1}}}.
\end{equation}

We note that there is a minor problem with preserving gauge invariance 
when calculating the graph (3h) in DRED. It is associated with an 
$\varepsilon$-dimensional part of one of the $n$-dimensional 
metric tensors $g_{\mu\nu}^n$ that arises in every partonic loop and 
hampers collecting together similar terms. However, 
this problem appears to be an artificial one, as in this particular case 
it makes no difference whether one uses 4- or $n$-dimensional metric 
tensor for the evaluation of this gluon self-energy graph. For this 
reason in practice one 
would set this $g_{\mu\nu}^n$ metric tensor to be the 4-dimensional one. 
Or, more exactly, if one introduces a proper counterterm so that to 
restore gauge invariance 
of the gluon self-energy, then the expression in DRED would be exactly the 
same as in (\ref{grh}), except for the term proportional to $N_C$, where 
one would have 28/9 instead of 31/9, leading to
\begin{equation}
\Delta(3h)=-B_s^{\mu\nu}.
\end{equation}

Concluding our discussion on the 2-point functions we remark that the 
matrix elements for 
the {\it additional u-channel 2-point functions} can be obtained from 
eqs.~(\ref{grd}), (\ref{grd1}) and (\ref{gre}) by 
the transformation (\ref{tu}).

We start by considering the t- and u-channel vertex diagrams. In 
particular we begin with the purely nonabelian graph (2b), which contains a 
four-point gluon vertex. It is finite, e.g. does not have UV and IR poles. 
For convenience we define the function
\begin{equation}
\label{lld}
f_{\rm lld} = 4 {\rm Li}_2(-x) + \ln^2(x) + 2 \zeta(2),
\end{equation}
with $\zeta(2) = \frac{\pi^2}{6}$.

Then the matrix element takes the form
\begin{eqnarray}
\nonumber
M_{\rm (2b)}^{\mu\nu} &=& i N_C \{ T^b T^a [ 4 (2 m g^{\mu\nu} - 2 
p_3^{\nu} 
\gamma^{\mu} - p_4^{\nu} \gamma^{\mu} + p_3^{\mu} \gamma^{\nu} + 2 p_4^{\mu} 
\gamma^{\nu}) (s \beta \ln(s/m^2) - m^2 f_{\rm lld}) -                 \\
\nonumber
&& 3 m s \gamma^{\mu}\gamma^{\nu} (4 \beta \ln(s/m^2) - f_{\rm lld}) ]
/4 s^2\beta^3    +  (a\leftrightarrow b, \mu\leftrightarrow\nu)  \}  - \\
\nonumber
&& i\delta^{ab} [ 2 (2 m g^{\mu\nu} + p_3^{\nu} \gamma^{\mu} - 
p_4^{\nu} \gamma^{\mu} + p_3^{\mu} \gamma^{\nu} - p_4^{\mu} \gamma^{\nu})
(s \beta \ln(s/m^2) - m^2 f_{\rm lld}) -                                \\
\label{b}
&& 3 m s \beta^2 g^{\mu\nu} f_{\rm lld} ]/4 s^2\beta^3.
\end{eqnarray}
It is easily seen from eq.~(\ref{b}) that graph (2b) is explicitly t-u 
symmetric, as it follows from its geometric topology.
Because $M_{\rm (2b)}^{\mu\nu}$ does not exibit any UV poles, there is no 
difference between DREG and DRED in this case, e.g.
\begin{equation}
\Delta(2b) = 0.
\end{equation}

Next we turn to graphs (2c1) and (2c2) and define another useful function:
\begin{equation}
\label{dl}
f_{\rm dl} = \zeta(2) - {\rm Li}_2(\frac{T}{m^2}).
\end{equation}
Diagrams of this topology do not only occur in hadroproduction, but also in 
other processes such as photoproduction and $\gamma\gamma$ production of 
heavy flavors. For this reason we also present the corresponding t-channel 
color factors for these graphs. 
Then it is 
straightforward to separate our Dirac structure from the color coefficients 
and one can easily deduce results for other processes where these graphs 
contribute, though with different color weights. The color factor for both 
(2c1) and (2c2) diagrams is the same:
\begin{equation}
T_{\rm col}^{\rm (2c1)} = T_{\rm col}^{\rm (2c2)} = 
(C_F - \frac{N_C}{2}) T^b T^a = - \frac{1}{6} T^b T^a.
\end{equation}
The complete matrix elements are:
\begin{eqnarray}
\nonumber
M_{\rm (2c1)}^{\mu\nu} &=& B_t^{\mu\nu} [ - 1/\varepsilon' 
           + 2 f_{\rm dl} m^2/t - \ln(-t/m^2) (6m^2+t)/T ]/6     \\
\label{c1}
&+&   i T^b T^a \{ p_3^{\mu} \gamma^{\nu}
[ 2 f_{\rm dl} m^2 T/t^2 + \ln(-t/m^2) (m^2/T - 2 T/t) + 1 ]   \\
\nonumber
&-& m p_3^{\mu} {\rm \hspace{-.05in}}   \not p_1 \gamma^{\nu}  
   [ \ln(-t/m^2)/T - 1/t ]  +
    m \gamma^{\mu} \gamma^{\nu} \ln(-t/m^2)    \}/3 T.
\end{eqnarray}
Because one has a UV pole $1/\varepsilon'$ (the 
prime denotes UV poles), there is a finite difference between 
the two n-dimensional schemes:
\begin{equation}
\Delta(2c1) = (C_F - \frac{N_C}{2}) B_t^{\mu\nu}.
\end{equation}
As expected, the difference is proportional to the sum of the Born terms of 
the relevant production channels. We emphasize
that one has to take care when calculating matrix elements in DRED.
In particular, one has to keep a clear distinction between the 
4-dimensional and the $n$-dimensional metric tensors $g^{\mu\nu}$ when 
doing their convolutions.
Failing to do so may very well result in introducing extra terms that 
do not satisfy the Slavnov-Taylor identities $[$\ref{Muta}$]$.

For the graph (2c2) we obtain:
\begin{eqnarray}
\nonumber
M_{\rm (2c2)}^{\mu\nu} &=& B_t^{\mu\nu} [ - 1/\varepsilon' 
           + 2 f_{\rm dl} m^2/t - \ln(-t/m^2) (6m^2+t)/T ]/6     \\
\label{c2}
&+&   i T^b T^a \{ p_4^{\nu} \gamma^{\mu}
[ - 2 f_{\rm dl} m^2 T/t^2 + \ln(-t/m^2) (m^2/T + 2 T/t) - 2m^2/t - 1 ] 
{\rm \hspace{.6in}}                             \\
\nonumber
&+& m p_4^{\nu} \gamma^{\mu}
{\rm \hspace{-.1in}}         \not p_1 [ \ln(-t/m^2)/T - 1/t ] +
    m \gamma^{\mu} \gamma^{\nu} \ln(-t/m^2)                    \\
\nonumber
&-& 2 m p_3^{\mu} p_4^{\nu} [ \ln(-t/m^2)/T - 1/t ]   \}/3 T
\end{eqnarray}
and
\begin{equation}
\Delta(2c2) = (C_F - \frac{N_C}{2}) B_t^{\mu\nu}.
\end{equation}

To write down the results for graphs (2c3) and (2c4) we introduce one 
more function:
\begin{equation}
\label{ll}
f_{\rm ll} = \ln^2(-\frac{t}{m^2}) + {\rm Li}_2(\frac{T}{m^2}).
\end{equation}
The color factors for both diagrams are the same:
\begin{equation}
T_{\rm col}^{\rm (2c3)} = T_{\rm col}^{\rm (2c4)} = 
- \frac{N_C}{2} T^b T^a = - \frac{3}{2} T^b T^a.
\end{equation}

Graphs (2c3) and (2c4) exhibit a rich structure of UV and IR singularities. 
Here one of the scalar integrals that is needed for the reduction of tensor 
integrals is of the type
\begin{equation}
\int \frac{d^n k}{k^2 (k+p)^2},
\end{equation}
which is effectively zero in dimensional regularization. However, we choose 
to separate UV and IR/M poles to keep trace of all the sources of UV 
singularities, e.g. the above mentioned integral is proportional to the 
difference $1/\varepsilon' - 1/\varepsilon$. This is our procedure for the 
relevant vertex diagrams. Thus,
\begin{eqnarray}
\nonumber
M_{\rm (2c3)}^{\mu\nu} &=& 3 B_t^{\mu\nu} [ 3/\varepsilon' - 
1/\varepsilon^2 + (2 \ln(-t/m^2) - 1)/\varepsilon 
+ 4 + 6 \ln(-t/m^2) m^2/T - 2 f_{\rm ll} ]/2        \\
\label{c3}
&+&  3 i T^b T^a \{  p_3^{\mu} \gamma^{\nu} [ -1/\varepsilon^2 + 
2/\varepsilon + 2 \ln(-t/m^2)/\varepsilon - 2 f_{\rm ll} +  
2 \ln(-t/m^2)                                          \\
\nonumber
&\times&     m^2 (m^2/T + 1)/T + 2 m^2/T + 6 ]/2 t 
\,\, - \,\, 3 m \gamma^{\mu}\gamma^{\nu} \ln(-t/m^2)/2 T    \\
\nonumber
&-&  m p_3^{\mu} {\rm \hspace{-.05in}} \not p_1 \gamma^{\nu} 
                     [ \ln(-t/m^2) (m^2+T) + T ]/t T^2     \} 
\end{eqnarray}
with
\begin{equation}
\Delta (2c3) = \frac{N_C}{2} B_t^{\mu\nu}.
\end{equation}
And
\begin{eqnarray}
\nonumber
M_{\rm (2c4)}^{\mu\nu} &=& 3 B_t^{\mu\nu} [ 3/\varepsilon' -
1/\varepsilon^2 + (2 \ln(-t/m^2) - 1)/\varepsilon 
+ 4 + 6 \ln(-t/m^2) m^2/T - 2 f_{\rm ll} ]/2        \\
\label{c4}
&+&  3 i T^b T^a \{  p_4^{\nu} \gamma^{\mu} [ 1/\varepsilon^2 -
2/\varepsilon - 2 \ln(-t/m^2)/\varepsilon + 2 f_{\rm ll} +
2 \ln(-t/m^2)                                          \\
\nonumber
&\times&     m^2 (m^2/T + 1)/T + 2 m^2/T - 6 ]/2 t
\,\, - \,\, 3 m \gamma^{\mu}\gamma^{\nu} \ln(-t/m^2)/2 T    \\
\nonumber
&+&  m p_4^{\nu}\gamma^{\mu}    {\rm \hspace{-.1in}}
\not p_1 [ \ln(-t/m^2) (m^2+T) + T ]/t T^2
- 2 m p_3^{\mu} p_4^{\nu} [ \ln(-t/m^2) (m^2+T) + T ]/t T^2  \} 
\end{eqnarray}
with
\begin{equation}
\Delta (2c4) =  \frac{N_C}{2} B_t^{\mu\nu}.
\end{equation}

The results for the matrix elements of the {\it additional u-channel vertex 
graphs} are obtained from eqs.~(\ref{c1}),(\ref{c2}),(\ref{c3}) and 
(\ref{c4}) by the transformation (\ref{tu}). However, there is a subtle 
point involved here: we stress that for the graphs (2c3) and (2c4) 
transformation (\ref{tu}) transforms the t-channel result of the graph 
(2c3) to the u-channel result for the graph (2c4), while the t-channel 
result of (2c4) goes to the u-channel result for (2c3). This is important 
to keep in mind when dealing with reactions which involve asymmetric set 
of graphs as e.g. photoproduction of heavy flavors.

It is worthwhile to mention that we have tested the above results for the 
t- and u-channel contributions against the ones given in $[$\ref{MCGb}$]$ 
and $[$\ref{KMC}$]$ 
by folding our NLO matrix elements with the Born term matrix elements 
for the cases of unpolarized and longitudinally polarized incoming bosons.
We obtain full agreement.

Next we turn to the remaining s-channel graphs shown in Fig.~3. 
For all the gluon propagators we work in Feynman gauge. 
Although this set of graphs is purely nonabelian for QCD type one-loop 
corrections, 
there could be also abelian (e.g. QED) virtual corrections to graph (3f1). 
For this reason we also give the color factor for it separately:
\begin{equation}
T_{\rm col}^{\rm (3f1)} = (C_F - \frac{N_C}{2}) ( T^a T^b - T^b T^a )
= - \frac{1}{6} ( T^a T^b - T^b T^a ).
\end{equation}
In the result for this graph together with the UV divergence we clearly see 
a collinear pole as well, multiplied by a logarithmic factor, that should 
be cancelled against the corresponding term in the factorization 
counterterm of the real bremsstrahlung: 
\begin{eqnarray}
\nonumber
M_{\rm (3f1)}^{\mu\nu} &=& \{  - B_s^{\mu\nu} s [ s\beta/\varepsilon' +
     (2m^2-s) ( 2\ln(x)/\varepsilon - 4 {\rm Li}_2(x) - 4 \ln(x)\ln(1 - x)
     + 6 \ln(x)                                    \\
&+& \ln^2(x) -  8 \zeta(2) ) + 3 \ln(x) s ]               \\
\nonumber
     &+& 2 i (T^a T^b - T^b T^a) m \ln(x) [ - g^{\mu\nu} (s + 2 t) - 
4 p_3^{\mu} p_4^{\nu} + 4 p_4^{\mu} p_3^{\nu}] \}/6 s^2\beta.
\end{eqnarray}
\begin{equation}
\Delta_{\rm (3f1)} =(C_F - \frac{N_C}{2}) B_s^{\mu\nu}.
\end{equation}

Graph (3f2) contributes as:
\begin{eqnarray}
\nonumber
M_{\rm (3f2)}^{\mu\nu} &=& N_C \{ B_s^{\mu\nu} [ 3 s \beta^2 
                                               (1/\varepsilon' + 2)
                 - \ln(s/m^2) (8 m^2 - s) + f_{\rm lld} m^2/\beta ]  \\
  &+& 2 i (T^a T^b - T^b T^a) m [ g^{\mu\nu} (s+2t) + 4 p_3^{\mu} p_4^{\nu} 
               - 4 p_4^{\mu} p_3^{\nu} ]                            \\
\nonumber
&\times&  [ \ln(s/m^2) (8 m^2 + s)/\beta^2 
               - 2s - 3 f_{\rm lld} m^2/\beta^3 ]/s^2 \}/2 s\beta^2.
\end{eqnarray}
The difference between the two regularization schemes is again determined 
by the coefficient of the UV pole:
\begin{equation}
\Delta_{\rm (3f2)} = \frac{N_C}{2} B_s^{\mu\nu} .
\end{equation}

We finish our consideration of the vertex diagrams for gluon fusion with 
the triangle graph contribution (tri)$\equiv$(3g1)+(3g2), e.g. we sum the 
two graphs (3g1) and (3g2).
For the case when one has gluons and ghosts inside the triangle loop we 
obtain:
\begin{eqnarray}
\nonumber
M_{\rm (tri)}^{\mu\nu}(g) &=& N_C \{ B_s^{\mu\nu} [ 33/\varepsilon' 
       - 36/\varepsilon^2 - 171/\varepsilon + 36\ln(s/m^2)/\varepsilon
       + 138 \ln(s/m^2) - 18 \ln^2(s/m^2)                   \\
\nonumber
&+&     144\zeta(2) - 284 ] 
 +  6 i (T^a T^b - T^b T^a) {\rm \hspace{-.1in}}  \not p_1
    [ g^{\mu\nu} ( 27/\varepsilon' - 6/\varepsilon^2 - 33/\varepsilon
      + 6 \ln(s/m^2)/\varepsilon                          \\
\label{glutri}
&+& 6 \ln(s/m^2) - 3 \ln^2(s/m^2)
      + 24 \zeta(2) - 4 )/s - p_2^{\mu} p_1^{\nu} 16/s^2 ]  \}/72;
\end{eqnarray}
For the two more cases when one has light and heavy quarks inside the 
loop we get
\begin{eqnarray}
\nonumber
M_{\rm (tri)}^{\mu\nu}(q) &=&  2 n_{lf} \{ B_s^{\mu\nu} [ 3/\varepsilon'
              - 3 \ln(s/m^2) + 5 ] - 3 i (T^a T^b - T^b T^a)
                                       {\rm \hspace{-.1in}}  \not p_1
                [ g^{\mu\nu}/s - 2 p_2^{\mu} p_1^{\nu}/s^2 ]   \}/9  \\
\label{qtri}
\end{eqnarray}
with $n_{lf}$ number of light flavors in the triangle loop, while for the 
heavy flavor case one has 
\begin{eqnarray}
\label{qmtri}
M_{\rm (tri)}^{\mu\nu}(Q) &=& 2  \{ B_s^{\mu\nu} [ 3/\varepsilon' +
         3 \ln(x) (2m^2/s+1)\beta + 5 + 12m^2/s ]                      \\
\nonumber
&-& 3 i (T^a T^b - T^b T^a)
                                       {\rm \hspace{-.1in}}  \not p_1
                [ g^{\mu\nu}/s^2 - 2 p_2^{\mu} p_1^{\nu}/s^3 ]
                [ 3 (\ln(x) + 4\beta)\ln(x) m^2                  \\
\nonumber
&-& 18\zeta(2) m^2 +
                24 m^2 + s ]                       \}   /9.
\end{eqnarray}
The complete matrix element for the triangle is the sum of the 
above three expressions (\ref{glutri}), (\ref{qtri}) and (\ref{qmtri}):
\begin{equation}
M_{\rm (tri)}^{\mu\nu} = M_{\rm (tri)}^{\mu\nu}(g) + 
                  M_{\rm (tri)}^{\mu\nu}(q) + M_{\rm (tri)}^{\mu\nu}(Q).
\end{equation}
We have compared our results in eqs.~(\ref{glutri}) and (\ref{qtri}) with 
those available in the literature $[$\ref{Andrei}$]$, and found agreement.

The difference between dimensional reduction and regularization arises 
due to gluons in the loop: 
\begin{equation}
\Delta_{\rm (tri)} = B_s^{\mu\nu}.
\end{equation}

\renewcommand{\theequation}{3.\arabic{equation}}
\setcounter{equation}{0}
\vglue 1cm
\begin{center}\begin{large}\begin{bf}
III. RESULTS FOR THE BOX DIAGRAMS IN GLUON FUSION
\end{bf}\end{large}\end{center}
\vglue .3cm

In this chapter we describe the technically most complicated derivation of 
the 4-point massive loop diagrams. The results of this chapter cannot be 
deduced 
from the results of any other relevant publications up to date.
The four box graphs (2a1)--(2a4) contributing to the subprocess 
$g+g\rightarrow Q+\overline Q$ are depicted 
in Fig.~2. We have used an adapted version of the Passarino-Veltman 
techniques $[$\ref{passar}$]$ to reduce tensor integrals to scalar ones. 
The scalar integrals are taken from $[$\ref{Been}$]$ which have been 
checked by us and in $[$\ref{Bojak1}$]$.

First of all we note that the results for the box diagrams are the same 
both in DREG and DRED, as a consequence of the ultraviolet convergence of 
the box graphs.

We expand all the box diagrams arising at the one loop level in gluon fusion
in terms of eight independent Dirac structures. In turn, coefficients
of the Dirac structures are expanded in terms of universal independent
Lorentz objects. And finally, the coefficients of the Lorentz structures are
expanded as products of a small set of analytic functions and various 
coefficient functions, which are combinations of scalar Mandelstam variables 
of the subprocess under consideration. Note that the Dirac structures and 
Lorentz objects are the same for every box diagram.
In particular, we cast the box matrix element into the following universal
form:
\begin{eqnarray}
\label{genbox}
M^{\mu\nu}&=&i T_{\rm col} \,\, \{ M_{\rm Bt}^{\mu\nu} \sum f_i  b_i   \\
\nonumber  &+& 
       \not{p}_1 [ g^{\mu\nu} \sum f_i  b_{i1}^{(p)} +
                   p_3^{\mu} p_3^{\nu} \sum f_i  b_{i2}^{(p)} +
                   p_3^{\mu} p_4^{\nu} \sum f_i  b_{i3}^{(p)} +
                   p_4^{\mu} p_3^{\nu} \sum f_i  b_{i4}^{(p)} +
                   p_4^{\mu} p_4^{\nu} \sum f_i  b_{i5}^{(p)} ]   \\
\nonumber  &+&
              \gamma^{\mu} [ p_3^{\nu} \sum f_i  b_{i1}^{(m)} +
                            p_4^{\nu} \sum f_i  b_{i2}^{(m)} ] +
              \gamma^{\nu} [ p_3^{\mu} \sum f_i  b_{i1}^{(n)} +
                            p_4^{\mu} \sum f_i  b_{i2}^{(n)} ] +
              m \gamma^{\mu} \gamma^{\nu} \sum f_i  c_i    \\
\nonumber  &+&
m \gamma^{\mu}{\rm\hspace{-.1in}}\not{p}_{1} [ p_3^{\nu} \sum f_i d_{i1} +
                                   p_4^{\nu} \sum f_i  d_{i2} ] +
m \gamma^{\nu}{\rm\hspace{-.1in}}\not{p}_{1} [ p_3^{\mu} \sum f_i e_{i1} +
                                   p_4^{\mu} \sum f_i  e_{i2} ]    \\
\nonumber  &+&
m [ g^{\mu\nu} \sum f_i  g_{i1} +
    p_3^{\mu} p_3^{\nu} \sum f_i  g_{i2} +
    p_3^{\mu} p_4^{\nu} \sum f_i  g_{i3} +
    p_4^{\mu} p_3^{\nu} \sum f_i  g_{i4} + 
    p_4^{\mu} p_4^{\nu} \sum f_i  g_{i5} ] \}   \\
\nonumber  &+&           {\cal M}_t \leftrightarrow {\cal M}_u.
\end{eqnarray}

Even though the number of independent covariants for the process 
$g+g\rightarrow Q+\overline Q$ is eight for $n=4$ this will not be the 
case for $n\neq 4$ relevant to this application. We have therefore made no 
attempt to reduce the above number of covariants to a minimal set.

Depending on the type of the box graph one has different number of terms
under the summation signs in (\ref{genbox}). These numbers as well as the 
analytic functions $f_i$ are specified below. The coefficient functions 
$b_i, ... , g_{i5}$ are given in Appendix~A of this paper.

The t-channel Born term matrix structure $M_{\rm Bt}^{\mu\nu}$ is the same 
for all box graphs and is defined as
\begin{equation}
M_{\rm Bt}^{\mu\nu} \equiv \gamma^{\mu} (\hat{p}_3-\hat{p}_1+m) \gamma^{\nu},
\end{equation}
which, when taken between the spin wave functions implying the 
effective relations $p_1^{\mu}=0,\,\, p_2^{\nu}=0$, can be written as 
\begin{equation}
\label{bt}
M_{\rm Bt}^{\mu\nu} = 2 p_3^{\mu} \gamma^{\nu}  -  \gamma^{\mu} 
           {\rm \hspace{-.1in}}               \not{p}_1 \gamma^{\nu}.
\end{equation}

Furthermore, for the box diagrams (2a1) and (2a2) we empirically found 
the following relations between the $d$ and $e$ coefficients:
\begin{equation}
\label{rel1}
e_{i1}=d_{i2},  {\rm \hspace{.6in}}   e_{i2}=d_{i1}.
\end{equation}

For the 4-point graph (2a4) the above relations are slightly changed: 
\begin{eqnarray}
\nonumber
e_{i1}&=&d_{i2},  {\rm \hspace{.4in}}   e_{i2}=d_{i1},
{\rm \hspace{.4in}  for  \hspace{.2in}}  i=1,2,4,5,8,9;    \\
\label{rel2}
e_{31} &=& d_{32} + 2 m^2 u/(m^2 s - u t)^2,  {\rm \hspace{.2in}}
e_{61} = 2 d_{62},   {\rm \hspace{.2in}}   e_{71} = 2 d_{72},     \\
\nonumber
e_{32} &=& d_{31} - 2 m^2 t/(m^2 s - u t)^2,  {\rm \hspace{.2in}}
e_{62} = 2 d_{61},   {\rm \hspace{.2in}}   e_{72} = 2 d_{71}.
\end{eqnarray}

Because of the relations (\ref{rel1}) and (\ref{rel2}), we will not write 
down the results for the $e$ coefficients in the Appendix~A.

We note that for the boxes (2a3) and (2a4) the ${\cal M}_t \leftrightarrow 
{\cal M}_u$ term in (\ref{genbox}) is {\it absent} as there does not exist 
a u-channel diagram for them. Furthermore, it is easy to see 
that boxes (2a3) and (2a4) go into each other with 
${\cal M}_t \leftrightarrow {\cal M}_u$. For this reason we write 
down explicit results only for one of these boxes in App.~A.

Next we present the color factors and functions for the abelian type box 
diagram (2a1). For this graph the sums over $i$ in (\ref{genbox}) run from 
1 to 7 for all the terms except for the term that multiplies 
$M_{\rm Bt}^{\mu\nu}$, which runs from 1 to 8. We have:
\begin{equation}
T_{\rm col}=\frac{1}{4} \delta^{ab} + (C_F - \frac{N_C}{2}) T^b T^a.
\end{equation}
\begin{eqnarray}
\label{fa1}
f_1 &=& \ln(x),  {\rm \hspace{.2in}}  f_2=\ln^2(x),          \\
\nonumber
f_3 &=& -2 \ln(x) \ln(1+x) + 2\ln (x) \ln(-\frac{t}{m^2}) - 
2 {\rm Li}_2(-x) + 3 \zeta(2),       {\rm \hspace{.2in}}
f_4 = {\rm f_{dl}},                                          \\
\nonumber
f_5 &=&\zeta(2), {\rm \hspace{.2in}}   f_6=\ln (-\frac{t}{m^2}),
{\rm \hspace{.2in}}     f_7=1,       {\rm \hspace{.2in}}
f_8 = \ln(x)\ln(1-x) + {\rm Li}_2(x).
\end{eqnarray}

For the nonabelian box diagram (2a2) the sums over $i$ in 
(\ref{genbox}) run from 1 to 8 for all terms except for the one that 
multiplies $M_{\rm Bt}^{\mu\nu}$, which runs from 1 to 9. For the color 
factor we obtain: 
\begin{equation}
T_{\rm col}=\frac{1}{4} \delta^{ab} + \frac{N_C}{2} T^b T^a.
\end{equation}
The relevant nine analytic functions that describe the result of 
evaluating the box diagram (2a2) are given by
\begin{eqnarray}
\nonumber
f_1 &=& \ln(\frac{s}{m^2}), {\rm \hspace{.2in}} f_2=\ln^2(\frac{s}{m^2}), 
{\rm \hspace{.2in}}
f_3 = \ln^2(\frac{s}{m^2}) + 4 {\rm f_{ll}} - 4 
\label{fa2}
\ln(\frac{s}{m^2}) \ln(-\frac{t}{m^2}) + 2 \zeta(2),          \\
f_4 &=& {\rm f_{lld}},   {\rm \hspace{.2in}}
f_5 =\zeta(2), {\rm \hspace{.2in}}   f_6=\ln (-\frac{t}{m^2}),
{\rm \hspace{.2in}}     f_7=1,  {\rm \hspace{.2in}}    f_8 = {\rm f_{ll}},
\\  \nonumber
f_9 &=& \ln(\frac{s}{m^2}) \ln (-\frac{t}{m^2}).
\end{eqnarray}

In the case of the crossed box (2a4) one has nine terms for every sum in 
(\ref{genbox}). The color factor for this graph takes the simple form
\begin{equation}
T_{\rm col}=\frac{1}{4} \delta^{ab}.
\end{equation}
We found it convenient to define functions $f_i$ as follows:
\begin{eqnarray}
\label{fa4}
f_1 &=& {\rm f_{ll}},                 {\rm \hspace{.2in}}
f_2 = {\rm f_{ll}} (t\rightarrow u),
\\
\nonumber
f_3 &=& {\rm f_{dl}} + {\rm f_{dl}} (t\rightarrow u) - 
        {\rm f_{ll}} - {\rm f_{ll}} (t\rightarrow u) + 
        2\ln(-\frac{t}{m^2})\ln(-\frac{u}{m^2}) - 4 \zeta(2),   \\
\nonumber
f_4 &=& {\rm f_{dl}},    {\rm \hspace{.2in}}
f_5 =\zeta(2), {\rm \hspace{.2in}}   f_6=\ln (-\frac{t}{m^2}),
{\rm \hspace{.2in}}     f_7=\ln (-\frac{u}{m^2}),               \\
\nonumber
f_8 &=& \ln(-\frac{t}{m^2})\ln(-\frac{u}{m^2}),   {\rm \hspace{.2in}}
f_9 = 1.
\end{eqnarray}

In order to compare with existing results we have folded the one-loop 
contribution with the LO Born term results.
Using our results for the box matrix elements we have reproduced all
the unpolarized and polarized amplitudes presented in the papers
$[$\ref{MCGb}$]$ and $[$\ref{KMC}$]$, where the Born term was 
folded in at the 
very beginning of the calculation. Note that after folding with the LO 
Born term many powers in the numerators and denominators cancel out, 
leading to a very short expressions for the box amplitudes.

However, the above checks and the ones mentioned in the Section~II did not 
include all the graphs for the gluon fusion subprocess. For a further 
check on the correctness of the results for all of our matrix 
elements in Figs.~2 and 3 we have performed a global check on the gauge 
invariance by first excluding one of the heavy quark momenta from the 
independent set of momenta (see Ref.~$[$\ref{Slaven}$]$ for details), 
e.g. reexpressing $p_4^{\nu}$ in terms of 
$p_1^{\nu}$ and $p_3^{\nu}$, then contracting the matrix elements by 
$p_1^{\mu}$. As well, we have performed the second possible check, 
contracting our matrix elements by the momentum of the other massless 
boson $p_2^{\nu}$. 
We have verified gauge 
invariance for the following gauge-invariant subsets of diagrams: (i) When 
the incoming gauge bosons are photons, i.e. including graphs (2a1), (2c1), 
(2c2), (2d1), (2d2), (2d3) and their u-channel counterparts, with 
corresponding color weights; (ii) For the photoproduction of heavy 
flavors, 
i.e. including all the above diagrams plus graphs (2a4), (2c4), (2e1) and 
their u-channel counterparts, with corresponding color weights; 
(iii) For the hadroproduction of heavy flavors, which ultimately includes 
all the graphs from Figs.~2 and 3 and their relevant u-channel 
counterparts.
We emphasize that the above gauge invariance checks were made separately 
for both color structures $C_F$ and $N_C$, and for every existing 
combination of color matrices $T^a$, $T^b$ and $\delta^{ab}$, whenever 
they arise. 
Also, for the hadroproduction set of graphs we had to set 
$\varepsilon'=\varepsilon$ as there is a mixing of UV and IR poles in the 
s-channel, particularly for the triangle graph (3g1) with gluons in the 
loop.

At the end of our discussion for the gluon fusion subprocess we comment on 
the DRED result. First we note that at the one-loop level the $Z_{1F}$ 
vertex renormalization 
constant is a sum of a wave function renormalization constant $Z_2$, which 
is equal to the abelian vertex renormalization constant $Z_{1V}$, and the 
term 
that is proportional to the $N_C$, which does not differ in DREG and DRED:
\begin{equation}
\label{z1f}
Z_{1F}=Z_2 - \frac{g^2}{\varepsilon} C_{\varepsilon}(\mu^2) N_C .
\end{equation}
Most importantly, the modification of the renormalization constants $Z_2$ 
and $Z_{1F}$ in DRED does not affect the other renormalization constants 
and preserves the form of the Slavnov-Taylor identities for all those 
constants by construction. 

Next we take the set of graphs relevant to heavy flavor production in 
photon-photon collisions which are the set of graphs mentioned in item 
(i) above. For this 
abelian set of graphs the overall action of the renormalization 
counterterms amounts to completely removing the quark self-energy 
graphs with on-shell legs and adding just one term that is equal to one of 
the removed graphs, e.g. expression (\ref{grd}) in DREG or 
(\ref{grddred}) in DRED. 
But this is effectively equivalent to multiplying each of the two 
self-energy graphs by a factor of 1/2, and leaving out renormalization. 
Then, correspondingly, the differences between the two dimensional schemes 
for these self-energy graphs have to be halved, and simple counting 
will now show that the sum of $\Delta$'s for this set of graph vanishes 
identically. 
Taking into account the above remark on the $Z_{1F}$ constant, e.g. 
separating the $C_F$ and $N_C$ parts of the vertex diagrams during 
the renormalization procedure, one can see that the same statements 
apply to the photoproduction set of graphs given by the 
gauge-invariant 
set (ii) above. And the same is true also for all the self-energy and 
vertex graphs for the hadroproduction set. The remaining 
difference coming from the gluonic loops in graphs (3g1) and (3h) cancel. 
This means that there is no difference between the DREG and DRED virtual 
NLO corrections for the gauge boson fusion subprocesses.

Finally we note that the original computer output for the box diagrams 
was extremely long. The final results were cast into the above 
shorter form 
with the help of the REDUCE Computer Algebra System $[$\ref{reduce}$]$.

\renewcommand{\theequation}{4.\arabic{equation}}
\setcounter{equation}{0}
\vglue 1cm
\begin{center}\begin{large}\begin{bf}
IV. ANNIHILATION OF THE QUARK-ANTIQUARK PAIR
\end{bf}\end{large}\end{center}
\vglue .3cm

The graphs contributing to this subprocess are shown in Fig.~4 for the 
leading order term and in Fig.~5 for the one-loop corrections. 
The leading order contribution proceeds only through the s-channel 
graph. One has:
\begin{equation}
B_{q\bar{q}} = i T^a T^a \bar{v}(p_2)\gamma^{\mu}u(p_1) 
                      \bar{u}(p_3)\gamma_{\mu}v(p_4)/s.
\end{equation}
Here the color matrices $T^a$ belong to different fermion lines that are 
connected by the gluon having color index $a$. We have again left out the 
factor $g^2$ in 
the above equation. In the Passarino-Veltman reduction for tensor 
integrals we use the same scalar integrals as those appearing in the gluon 
fusion subprocess, with relevant shifts and interchanges of momenta as 
needed. 

Starting again with the 2-point functions, we notice that the result for 
graph (5g) can be 
obtained from the one of (\ref{grh}) for graph (2h) in the gluon fusion 
subprocess by the simple replacement
\begin{equation}
M_{(5g)} = M_{(2h)}^{\mu\nu} \, (B_s^{\mu\nu} \rightarrow B_{q\bar{q}}),
\end{equation}
and all the statements after (\ref{grh}) are equally applicable to 
$M_{(5g)}$.

The massless quark self-energy graphs (5j) and (5k) with external legs 
on-shell vanish identically:
\begin{equation}
M_{(5j)} = M_{(5k)} = 0.
\end{equation}

The massive quark self-energy graphs (5h) and (5i) with external 
legs on-shell are derived analogously to the ones considered in the 
previous section:
\begin{equation}
M_{\rm (5h)} = M_{\rm (5i)} = - C_F B_{q\bar{q}}
           \left( \frac{1}{\varepsilon'} + 4 + \frac{2}{\varepsilon} 
\right),
\end{equation}
and the difference between the two regularizations schemes is
\begin{equation}
\Delta(5h) = \Delta(5i) = - C_F B_{q\bar{q}}.
\end{equation}

Results for the vertex diagrams are relatively short. Starting with graphs 
(5c) and (5d) one finds that they are proportional to the LO Born term:
\begin{equation}
M_{(5c)} = B_{q\bar{q}} [ -1/\varepsilon' + 2/\varepsilon^2 + 
               4/\varepsilon - 2 \ln(s/m^2)/\varepsilon + \ln^2(s/m^2) - 3 
               \ln(s/m^2) - 8 \zeta(2) + 8 ]/6
\end{equation}
and
\begin{equation}
M_{(5d)} = 3 B_{q\bar{q}} [ 3/\varepsilon' - 4/\varepsilon + 
                 \ln(s/m^2) - 2 ]/2.
\end{equation}
Corresponding differences between DRED and DREG results are
\begin{equation}
\Delta (5c) = (C_F - \frac{N_C}{2}) B_{q\bar{q}},   {\rm \hspace{.6in}}
\Delta (5d) = \frac{N_C}{2} B_{q\bar{q}} .
\end{equation}

For the other two vertex diagrams we also obtain simple expressions:
\begin{eqnarray}
\nonumber
M_{(5e)} &=&  \{  B_{q\bar{q}} [ - 1/\varepsilon' + 
                  3\ln(x)\beta + (1/\beta+\beta) (\ln(x)/\varepsilon - 
                  2 {\rm Li}_2(x) - 2\ln(x)\ln(1 - x)                 \\
&+& \ln^2(x)/2 - 4\zeta(2)) ] + 4 i T^a T^a m 
\bar{v}(p_2) {\rm \hspace{-.1in}} \not p_3 u(p_1)    \bar{u}(p_3) v(p_4)
                                 \ln(x)/s^2\beta      \}/6
\end{eqnarray}
and
\begin{eqnarray}
M_{(5f)} &=&  3 \{  B_{q\bar{q}} [ 3 (1/\varepsilon' + 2) -
          \ln(s/m^2) (8 m^2/s - 1)/\beta^2 + {\rm f_{lld}} m^2/s\beta^3 ] 
           - 4 i T^a T^a m                                   \\
\nonumber   &\times&
\bar{v}(p_2) {\rm \hspace{-.1in}} \not p_3 u(p_1)   \bar{u}(p_3) v(p_4)
                          [ \ln(s/m^2) (8 m^2/s + 1)/\beta^2 -
                      2 - 3 {\rm f_{lld}} m^2/s\beta^3 ]/s^2\beta^2  \}/2
\end{eqnarray}
with the function ${\rm f_{lld}}$ defined in (\ref {lld}).
Once again, the differences between the two regularization schemes are
\begin{equation}
\Delta (5e) = (C_F - \frac{N_C}{2}) B_{q\bar{q}},   {\rm \hspace{.6in}}
\Delta (5f) = \frac{N_C}{2} B_{q\bar{q}} .
\end{equation}

Turning to the two box diagrams we again note that because of the absence 
of UV poles there are no differences between DRED and DREG. Extensive 
Dirac algebra manipulations lead to rather compact expressions for the 
matrix elements. 

Since for the $q\bar{q}\rightarrow Q\overline Q$ subprocess one has two
spinor ``sandwiches'' 
we cannot have momenta with Lorentz indices, and consequently there 
is no expansion of matrix elements in terms of Lorentz objects.
We expanded the box diagrams in terms of the seven independent Dirac 
structures, the same set for each of the two box graphs.
Then every Dirac structure is multiplied by the sums of products of a 
small set of analytic functions and coefficient functions. 
Thus, we have the following compact expansion for both box diagrams:
\begin{eqnarray}
\label{boxq}
M&=&i T_{\rm col} \,\, \{ 
\bar{v}(p_2) \gamma^{\mu} u(p_1) \bar{u}(p_3) \gamma_{\mu} v(p_4)
                                                     \sum f_i  h_i  \\ 
\nonumber  &+&
\bar{v}(p_2) {\rm \hspace{-.1in}} \not p_3 u(p_1)
\bar{u}(p_3) {\rm \hspace{-.1in}} \not p_1 v(p_4)
                                              \sum f_i h_{i}^{(1)}   \\
\nonumber  &+&
\bar{v}(p_2)\gamma^{\nu}{\rm \hspace{-.1in}}\not p_3\gamma^{\mu} u(p_1)
\bar{u}(p_3)\gamma_{\mu}{\rm \hspace{-.05in}}\not p_1\gamma_{\nu} v(p_4)
                                              \sum f_i h_{i}^{(2)}   \\
\nonumber  &+&
\bar{v}(p_2)\gamma^{\nu}\gamma^{\alpha}\gamma^{\mu} u(p_1)
\bar{u}(p_3)\gamma_{\mu}\gamma_{\alpha}\gamma_{\nu} v(p_4)
                                              \sum f_i h_{i}^{(3)}   \\
\nonumber  &+&
m \bar{v}(p_2) {\rm \hspace{-.1in}} \not p_3 u(p_1)
  \bar{u}(p_3) v(p_4)
                                              \sum f_i h_{i}^{(4)}   \\
\nonumber  &+&
m \bar{v}(p_2)\gamma^{\mu} u(p_1)
\bar{u}(p_3)\gamma_{\mu}{\rm \hspace{-.05in}}\not p_1 v(p_4)
                                              \sum f_i h_{i}^{(5)}   \\
\nonumber  &+&
m \bar{v}(p_2)\gamma^{\nu}{\rm \hspace{-.1in}}\not p_3\gamma^{\mu} u(p_1)
\bar{u}(p_3)\gamma_{\mu}\gamma_{\nu} v(p_4)
                                              \sum f_i h_{i}^{(6)}  \}.
\end{eqnarray}
Note that the number of independent covariants in $n\neq 4$ exceeds the 
number of independent covariants in $n=4$ where one has four independent 
covariants.

The sums over $i$ in (\ref{boxq}) run from 1 to 4 except for 
the first term (it is proportional to the Born term), where the sum runs 
from 1 to 6. Below we list the color factors and analytic functions for the 
two 4-point functions of (\ref{boxq}). For the graph (5a) we get:
\begin{equation}
\label{colqa}
T_{\rm col} = (T^a T^b) (T^b T^a),
\end{equation}
where the first parentheses in (\ref{colqa}) corresponds to the summation 
over color indices of the massless fermion line and 
\begin{eqnarray}
\label{fqa}
f_1 &=& \ln^2(\frac{s}{m^2}) + 4 {\rm f_{ll}} - 4
                   \ln(\frac{s}{m^2}) \ln(-\frac{t}{m^2}) + 2 \zeta(2),
\\      \nonumber
f_2 &=& \ln(-\frac{t}{m^2}),     {\rm \hspace{.3in}}
f_3 = \ln(\frac{s}{m^2}),      {\rm \hspace{.3in}}
f_4 = {\rm f_{lld}},         {\rm \hspace{.3in}}
f_5 = {\rm f_{ll}},         {\rm \hspace{.3in}}
f_6 = 1.
\end{eqnarray}
The color factor and the corresponding functions for the second box graph 
(5b) are
\begin{equation}
\label{colqb}
T_{\rm col} = (T^a T^b) (T^a T^b),
\end{equation}
and all the functions are obtained from the ones in (\ref{fqa}) by 
the simple interchange $t\rightarrow u$, e.g.:
\begin{eqnarray}
\label{fqb}
f_1 &=& \ln^2(\frac{s}{m^2}) + 4 {\rm f_{ll}}(t\rightarrow u) - 4
                   \ln(\frac{s}{m^2}) \ln(-\frac{u}{m^2}) + 2 \zeta(2),
\\      \nonumber
f_2 &=& \ln(-\frac{u}{m^2}),     {\rm \hspace{.3in}}
f_3 = \ln(\frac{s}{m^2}),      {\rm \hspace{.3in}}
f_4 = {\rm f_{lld}},         {\rm \hspace{.3in}}
f_5 = {\rm f_{ll}}(t\rightarrow u),         {\rm \hspace{.3in}}
f_6 = 1.
\end{eqnarray}

The coefficients $h_i, h_i^{(j)}$ are given in Appendix~B of this paper. 
However, there exists a partial symmetry for these box diagrams, which 
allows one to 
express most coefficients for the box graph (5b) through the ones of the 
box graph (5a). In particular, starting from the coefficients $h_i^{(j)}$ 
with superscript $j\geq 2$, we find the following general relations for 
all these coefficients:
\begin{eqnarray}
\label{relsq}
h_i^{(j)} [{\rm (5b)}] &=& - h_i^{(j)} [{\rm (5a)}] (t\rightarrow u),
{\rm \hspace{.3in}}  j=2,4;  \,\,\,  i=1\div 4;                    \\
\nonumber
h_i^{(j)} [{\rm (5b)}] &=& h_i^{(j)} [{\rm (5a)}] (t\rightarrow u),
{\rm \hspace{.3in}}  j=3,5,6;  \,\,\,  i=1\div 4.
\end{eqnarray}
Consequently, for the graph (5b) only the coefficients $h_i$ and 
$h_i^{(1)}$ 
are presented in Appendix~B. We should also mention that all the one-loop 
matrix elements of this chapter must be multiplied by the common factor 
(\ref{common}).

At the end of this section we must again say a few words about the DRED 
result. 
In line of the discussion in the previous section one has a cancellation 
of differences 
between DREG and DRED for the heavy quark loops: the two massive 
self-energy 
diagrams (5h) and (5i) versus the heavy quark vertex graphs (5e) and (5f). 
Concerning the differences arising from the two massless vertex diagrams 
(5c),(5d) and the off-shell gluon self-energy (5g), these differences 
remain, leading to an overall difference
\begin{equation}
\Delta(q\bar{q}\rightarrow Q\overline {Q}) = B_{q\bar{q}}/3 .
\end{equation}
First we stress that as the above difference is proportional to the 
LO Born term, it is manifestly gauge invariant. Secondly, 
it can be considered as a conversion term between DREG and DRED for the 
virtual corrections for this subprocess.

%In principle, we could have
%chosen the {\it massless} vertex renormalization constant in DRED so that
%it would take care of the differences for massless vertex diagrams. E.g.
%this would be the case if one defines the $\overline {\rm MS}$ massless
%wave function renormalization constant in DRED as
%\begin{equation}
%Z_2 = 1 - g^2 C_F C_{\varepsilon}(\mu^2) \left(
%                              \frac{1}{\varepsilon'} + 1 \right).     
%\end{equation}
%In the above equation we have chosen the $\overline {\rm MS}$ $Z_2$
%so as to cancel differences for the off-shell massless quark self-energy
%diagram that arise when one evaluates it in the two different
%regularization schemes. In this case the only difference in DRED comes
%from the gluon self-energy (5g).

\renewcommand{\theequation}{5.\arabic{equation}}
\setcounter{equation}{0}
\vglue 1cm
\begin{center}\begin{large}\begin{bf}
V. ABSORPTIVE PARTS 
\end{bf}\end{large}\end{center}
\vglue .3cm

In the previous sections we have been writing down expressions 
for the real parts of the matrix elements. However, for some physical 
applications as e.g. the calculation of $T$-odd observables the imaginary 
(or absorptive) parts of the one-loop amplitudes are needed as well. 
In addition, the NNLO calculation of heavy quark production involves also 
the product of the NLO one-loop contributions including their imaginary 
parts.

In order to derive the imaginary parts of our matrix elements we have 
proceeded in two 
different ways. Firstly, we applied Cutkosky rules $[$\ref{Cutkosky}$]$ to 
the scalar integrals that enter the Passarino-Veltman decomposition for our 
tensor integrals and calculated their imaginary parts directly, as 
discontinuities across the unitarity cut in the physical s-channel of the 
corresponding Feynman diagram. 
The result of these considerations may be cast into the form of an 
effective rule which states that the imaginary parts of our scalar 
integrals, and consequently the full 
matrix elements including real and imaginary pieces, can be 
obtained with the four simple substitutions in all of our expressions for 
matrix elements:
\begin{eqnarray}
\label{imag}
\ln(\frac{s}{m^2}) &\rightarrow& \ln(\frac{s}{m^2}) - i\pi,
{\rm \hspace{1.3in}}
\ln(x) \rightarrow \ln(x) + i\pi,                      \\
\nonumber
\ln^2(\frac{s}{m^2}) &\rightarrow& \ln^2(\frac{s}{m^2}) - 2 i\pi 
\ln(\frac{s}{m^2}),
{\rm \hspace{.6in}}
\ln^2(x) \rightarrow \ln^2(x) + 2 i\pi \ln(x).
\end{eqnarray}
Note that the other logarithmic and dilog functions do not possess any 
imaginary parts.
Secondly, to double-check our prescription (\ref{imag}), we have 
carefully traced the sign of the causal term +$i0$ in the propagators of 
the scalar integrals during the whole course of their derivation. We came 
to the conclusion that all the 
analytic functions are well defined in the physical region except for the 
two logarithms, which have negative arguments. They enter the expressions 
for the scalar integrals as $\ln(-\frac{s+i0}{m^2})$ and $\ln(-x+i0)$.
In other words, in all the expressions for our amplitudes the above two 
logarithms would actually appear (instead of $\ln(s/m^2)$ and 
$\ln(x)$), if we had not already extracted $-\pi^2$ terms from 
$\ln^2(-s/m^2-i0)$ and $\ln^2(-x+i0)$ and considered them in the real 
parts of the amplitudes, and this is fully consistent with (\ref{imag}). 

In case one desires to perform crossing and obtain results for other 
subprocesses many of the analytic functions of (\ref{fa1}), (\ref{fa2}), 
(\ref{fa4}), (\ref{fqa}) and (\ref{fqb}) will develop imaginary parts. The 
rules for deriving such 
imaginary contributions were considered in detail in Ref.~$[$\ref{KMM}$]$.
To apply those rules one needs to know the signs of the causal terms 
in our analytic functions. Thus, the only remaining thing to write down is 
a relative sign between the parameters of the analytic functions and the 
causal term $i0$ for the case of our kinematics:
\begin{equation}
s \rightarrow s + i0,    {\rm \hspace{.4in}}   t \rightarrow t + i0,
{\rm \hspace{.4in}}   u \rightarrow u + i0,    {\rm \hspace{.4in}} 
x \rightarrow x - i0.
\end{equation}

\renewcommand{\theequation}{9.\arabic{equation}}
\setcounter{equation}{0}
\vglue 1cm
\begin{center}\begin{large}\begin{bf}
VI. CONCLUSIONS
\end{bf}\end{large}\end{center}
\vglue .3cm

In this paper we have presented complete analytic results for the
one-loop contributions to heavy flavor production in a closed form, 
including their absorptive parts. 
These include the one-loop matrix elements of the relevant partonic
subprocesses (\ref{gluglu}) and (\ref{qbarq}) that are presented here for 
the first time$^2$. We have also indicated the way of deriving matrix 
elements 
for the other processes that can be obtained from those presented in this 
paper by crossing to different production channels. Our results 
are relevant not only for various NLO applications, but produce that part 
of the next-to-next-to-leading order corrections to heavy flavor 
production corresponding to the 
square of the one-loop matrix elements. Of course, to 
conclude the latter task the {\cal O}($\varepsilon^2$) expansion of the 
relevant NLO matrix element is also required. We reserve this task for a 
future publication.
\addtocounter{footnote}{1}
\footnotetext{We would be happy to provide our one-loop results in REDUCE 
format. Please contact Z.M. by e-mail.}

\vglue 1cm
\begin{center}\begin{large}\begin{bf}
ACKNOWLEDGEMENTS
\end{bf}\end{large}\end{center}
\vglue .3cm
We are grateful to A.~Davydychev and A.~Pivovarov for very informative
discussions.
Z.~M. would like to thank the Particle Theory group of the Institut
f{\"u}r Physik, Universit{\"a}t Mainz, and the Physics Department of McGill
University for hospitality. The work of Z.~M. was supported by a DFG grant
under contract 436 GEO 17/3/02.

\vglue 1cm
\begin{center}\begin{large}\begin{bf}
APPENDIX A
\end{bf}\end{large}\end{center}
\vglue .3cm

\setcounter{equation}{0}
\renewcommand{\theequation}{A\arabic{equation}}

Here we present the coefficients of the box contributions for the gluon 
fusion subprocess appearing in eq.~(\ref{genbox}).

Let us introduce the notation:
\begin{eqnarray}
\nonumber
z_1 &\equiv& m^2 s - t^2,   {\rm \hspace{.4in}}    z_2 \equiv s + 2 t,
{\rm \hspace{.4in}}
z_t \equiv 2 m^2 + t,   {\rm \hspace{.4in}}     z_u \equiv 2 m^2 + u,   \\
D &\equiv& m^2 s - u t.
\end{eqnarray}

First we list coefficients for the abelian type of box diagram (2a1):
\begin{eqnarray}
\nonumber
b_1 &=& (1/\beta + \beta)/\varepsilon t,   {\rm \hspace{.2in}}
b_2 = (z_u - u \beta)/2 D,  {\rm \hspace{.2in}}
b_3 = z_t z_u/t \beta D,    {\rm \hspace{.2in}}
b_4 = 2 z_u/D,            {\rm \hspace{.2in}}         \\
\nonumber
b_5 &=& (4 u \beta - 3 z_u)/D, {\rm \hspace{.2in}}  b_6=b_7=0,
{\rm \hspace{.2in}}
b_8 = -2 (1/\beta + \beta)/t;              \\
\nonumber
b_{11}^{(p)} &=& 2 z_t \beta/D,  {\rm \hspace{.2in}}
b_{21}^{(p)} = (b_{01}^{(p)} + b_{31}^{(p)})/2,   {\rm \hspace{.2in}}
b_{31}^{(p)} = s t^2 \beta^3/D^2,                      ß\\
\nonumber
b_{41}^{(p)} &=& -2 z_t (2 m^2/t - s t \beta^2/D)/D,  {\rm \hspace{.2in}}
b_{51}^{(p)} = -3 b_{01}^{(p)} - 4 b_{31}^{(p)},  {\rm \hspace{.2in}}
b_{61}^{(p)} = 2 z_t^2/T D,   {\rm \hspace{.2in}}
b_{71}^{(p)} = 0,                                      \\
\nonumber
&& {\rm with  \hspace{.2in}}
b_{01}^{(p)} = (z_t D + t z_1 \beta^2)/D^2;           \\
\nonumber
b_{12}^{(p)} &=& -4 m^2 \beta (z_t + z_1/s)/D^2,  {\rm \hspace{.2in}}
b_{22}^{(p)} = (b_{02}^{(p)} + b_{32}^{(p)})/2,          \\
\nonumber
b_{32}^{(p)} &=& 4 m^2 t \beta (D (2+t/s) - s t \beta^2)/D^3,
{\rm \hspace{.2in}}
b_{42}^{(p)} = -8 m^2 (s \beta^2 (T D + t^2 z_t) + 
                          m^2 z_2 T D/t)/t D^3,        \\
\nonumber
b_{52}^{(p)} &=& -3 b_{02}^{(p)} - 4 b_{32}^{(p)},  {\rm \hspace{.2in}}
b_{62}^{(p)} = 4 m^2 (2 D/t + s \beta^2 - 4 t u/s - t^2/T)/D^2,
{\rm \hspace{.2in}}     b_{72}^{(p)} = 4 z_t u/s t D,  \\
\nonumber
&& {\rm with  \hspace{.2in}}
b_{02}^{(p)} = 4 m^2 t ( - z_1 \beta^2 + u z_2 D/s^2)/D^3;   \\
\nonumber
b_{13}^{(p)} &=& -4 \beta ( 2 (2 m^2-s) z_t D/s^2 \beta^2 -
             2 m^2 s \beta^2 - 3 (m^2/s+1) D + 2 m^2 t z_2/s )/D^2,  \\
\nonumber
b_{23}^{(p)} &=& (b_{03}^{(p)} + b_{33}^{(p)})/2,   {\rm \hspace{.2in}}
b_{33}^{(p)} = -2 z_t (t (\beta+2 m^2 z_2/s^2\beta) D + 
                                              2 m^2 u z_2 \beta)/D^3, \\
\nonumber
b_{43}^{(p)} &=& -4 (2 m^4 z_2 s \beta^2 + 
        (4 m^2 T D/t^2 - z_t (2 m^4/t - t) + 2 m^2 z_2) D)/D^3,   \\
\nonumber
b_{53}^{(p)} &=& -3 b_{03}^{(p)} - 4 b_{33}^{(p)},  {\rm \hspace{.2in}}
b_{63}^{(p)} = 4 z_t (2 (D/t - z_1/s + 2 z_2) + 3 t u/T - t^2 (m^2 + 2 
                                                            t)/T^2)/D^2, \\
\nonumber
b_{73}^{(p)} &=& 4 z_t (u/s - m^2/T)/t D,                       \\
\nonumber
&& {\rm with  \hspace{.2in}}
b_{03}^{(p)} = 2 (2 m^2 u z_1 \beta^2 - (2 m^2 T - 2 m^2 u (u^2 + t^2)/s^2 
                                         + t^2 \beta^2) D)/D^3;    \\
\nonumber
b_{14}^{(p)} &=& -4 \beta (z_t D/s\beta^2 + 3 T^2 - t^2 (m^2 - t)/s)/D^2,
{\rm \hspace{.2in}}
b_{24}^{(p)} = (b_{04}^{(p)} + b_{34}^{(p)})/2,                \\
\nonumber
b_{34}^{(p)} &=& 2 t (4 m^2 D^2/s^2\beta + 2 m^2 u \beta D/s - 
                       t (3 s T + t^2) \beta^3)/D^3,                \\
\nonumber
b_{44}^{(p)} &=& 4 z_t ((2 m^4/t - t) D - 2 s t T \beta^2)/D^3,     \\
\nonumber
b_{54}^{(p)} &=& -3 b_{04}^{(p)} - 4 b_{34}^{(p)},  {\rm \hspace{.2in}}
b_{64}^{(p)} = -8 (z_t/s + t^2 \beta^2/D)/D,   {\rm \hspace{.2in}}
b_{74}^{(p)} = -4 z_t/s D,                     \\
\nonumber
&& {\rm with  \hspace{.2in}}
b_{04}^{(p)} = 2 (2 t (m^2 t^2 - s T^2) \beta^2 - (6 m^2 D/s + t z_t
                            + 2 m^2 t^2 z_2/s^2) D)/D^3;      \\
\nonumber
b_{15}^{(p)} &=& -4 m^2 \beta (z_t + z_1/s)/D^2,   {\rm \hspace{.2in}}
b_{25}^{(p)} = (b_{05}^{(p)} + b_{35}^{(p)})/2,                 \\
\nonumber
b_{35}^{(p)} &=& 4 m^2 t \beta ((2 + t/s) D - s t \beta^2)/D^3,
{\rm \hspace{.2in}}
b_{45}^{(p)} = -8 m^2 (T (D - t z_t) D/t^2 + s t z_t \beta^2)/D^3,  \\
\nonumber
b_{55}^{(p)} &=& -3 b_{05}^{(p)} - 4 b_{35}^{(p)},  {\rm \hspace{.2in}}
b_{65}^{(p)} = 4 m^2 (2 D/t + s \beta^2 - 4 t u/s - t^2/T)/D^2,
{\rm \hspace{.2in}}
b_{75}^{(p)} = 4 z_t u/s t D,            \\
\nonumber
&& {\rm with  \hspace{.2in}}
b_{05}^{(p)} = 4 m^2 t (u z_2 D/s^2 - z_1 \beta^2)/D^3;    \\
\nonumber
b_{11}^{(m)} &=& -2 (4 m^2/s - 3 + t z_t/D)/s\beta,  {\rm \hspace{.2in}}
b_{21}^{(m)} = (b_{01}^{(m)} + b_{31}^{(m)})/2,           \\
\nonumber
b_{31}^{(m)} &=& t (4 m^2 u/s^2\beta + (2 s T + t^2) \beta/D)/D,
{\rm \hspace{.2in}}
b_{41}^{(m)} = -2 (2 m^2 T/t - z_t (2 s T + t^2)/D)/D,    \\
\nonumber
b_{51}^{(m)} &=& -3 b_{01}^{(m)} - 4 b_{31}^{(m)},  {\rm \hspace{.2in}}
b_{61}^{(m)} = 2 (z_t - 2 t u/s)/D,               {\rm \hspace{.2in}}
b_{71}^{(m)} = 0,                            \\
\nonumber
&& {\rm with  \hspace{.2in}}
b_{01}^{(m)} = (2 (T - m^2 u/s) - t^2 z_t/D)/D;    \\
\nonumber
b_{12}^{(m)} &=& 2 (\beta - t z_u/\beta D)/s,     {\rm \hspace{.2in}}
b_{22}^{(m)} = (b_{02}^{(m)} + b_{32}^{(m)})/2,           \\
\nonumber
b_{32}^{(m)} &=& -t^2 (4 m^2/s^2\beta - u \beta/D)/D,   {\rm \hspace{.2in}}
b_{42}^{(m)}  = -2 (2 m^2 (m^2/t + 2) - t u z_t/D)/D,     \\
\nonumber
b_{52}^{(m)} &=& -3 b_{02}^{(m)} - 4 b_{32}^{(m)},      {\rm \hspace{.2in}}
b_{62}^{(m)} = 2 (2 D/s + 2 s + t + t^2/T)/D,        {\rm \hspace{.2in}}
b_{72}^{(m)} = 0,                                       \\
\nonumber
&& {\rm with  \hspace{.2in}}
b_{02}^{(m)} = (- 2 m^2 (1-t/s) + t u z_t/D)/D;           \\
\nonumber
b_{11}^{(n)} &=& -2 (2 - 2 m^2 t z_2/s D + 3 m^2 s \beta^2/D)/s\beta,
{\rm \hspace{.2in}}
b_{21}^{(n)} = (b_{01}^{(n)} + b_{31}^{(n)})/2,           \\
\nonumber
b_{31}^{(n)} &=& (2 (T + 2 m^2 t^2/s^2)/\beta - (2 m^2 z_1 - t^2 u) \beta/D)/D,
\\    \nonumber
b_{41}^{(n)} &=& 2 (2 m^2 (m^2/t + 2) D - 2 m^2 s t \beta^2 - t^2 z_u)/D^2,
{\rm \hspace{.2in}}
b_{51}^{(n)} = -3 b_{01}^{(n)} - 4 b_{31}^{(n)},      \\
\nonumber
b_{61}^{(n)} &=& -2 (6 m^2 + 2 u^2/s - 5 m^2 t/T)/D,     {\rm \hspace{.2in}}
b_{71}^{(n)} = 0,                                   \\
\nonumber
&& {\rm with  \hspace{.2in}}
b_{01}^{(n)} = (2 m^2 (1-t/s) - 2 m^2 s t \beta^2/D - t^2 z_u/D)/D;  \\
\nonumber
b_{12}^{(n)} &=& -2 (2 t/s - 2 t^2 z_t/s D + 3 s T \beta^2/D)/s\beta,
{\rm \hspace{.2in}}
b_{22}^{(n)} = (b_{02}^{(n)} + b_{32}^{(n)})/2,           \\
\nonumber
b_{32}^{(n)} &=& (2 (-2 m^2 t u/s^2 - T)/\beta - (2 m^2 s T - t^2 z_t)\beta/D)
                                                           /D,
\\    \nonumber
b_{42}^{(n)} &=& 2 (2 m^2 T D/t - 2 s t T \beta^2 - t^2 z_t)/D^2,
{\rm \hspace{.2in}}
b_{52}^{(n)} = -3 b_{02}^{(n)} - 4 b_{32}^{(n)},      \\
b_{62}^{(n)} &=& -2 (3 z_t + 2 t^2/s)/D,     {\rm \hspace{.2in}}
b_{72}^{(n)} = 0,                                   \\
\nonumber  
&& {\rm with  \hspace{.2in}}
b_{02}^{(n)} = t (2 m^2/s - 1 - z_1 \beta^2/D + t z_u/D)/D;     \\
\nonumber
c_1 &=& 4/s\beta,  {\rm \hspace{.2in}}   c_2 = (c_3 + c_4/2)/2,
{\rm \hspace{.2in}}   c_3 = -(s \beta + z_t/\beta + 2 t z_2/s\beta)/D, \\
\nonumber
c_4 &=& 2 (s + 3 t)/D,       {\rm \hspace{.2in}}
c_5 = - 4 c_3 - 3 c_4/2,    {\rm \hspace{.2in}}   c_6=c_7=0;    \\
\nonumber
d_{11} &=& 4 z_t/s\beta D,       {\rm \hspace{.2in}}
d_{21} = (d_{31} + d_{41}/2)/2,    {\rm \hspace{.2in}}
d_{31} = -2 T (2/s\beta + s \beta/D)/D,                         \\
\nonumber
d_{41} &=& 4 T z_2/D^2,    {\rm \hspace{.2in}}
d_{51} = - 4 d_{31} - 3 d_{41}/2,    {\rm \hspace{.2in}}
d_{61} = -4/D,       {\rm \hspace{.2in}}    d_{71} = 0;  \\
\nonumber
d_{12} &=& 4 z_u/s\beta D,       {\rm \hspace{.2in}}
d_{22} = (d_{32} + d_{42}/2)/2,    {\rm \hspace{.2in}}
d_{32} = 2 (2 m^2 z_1 - t u z_2)/s\beta D^2,                         \\
\nonumber
d_{42} &=& 4 t z_u/D^2,    {\rm \hspace{.2in}}
d_{52} = - 4 d_{32} - 3 d_{42}/2,    {\rm \hspace{.2in}}
d_{62} = -4 m^2/T D,       {\rm \hspace{.2in}}    d_{72} = 0;  \\
\nonumber
g_{11} &=& 2 (t\beta/D - 4/s\beta),   {\rm \hspace{.2in}}
g_{21} = (g_{31} + g_{41}/2)/2,   {\rm \hspace{.2in}}
g_{31} = t (4 z_t D/s\beta - t z_2 \beta)/D^2,         \\
\nonumber
g_{41} &=& -2 (6 T D - s t^2 \beta^2)/D^2,   {\rm \hspace{.2in}}
g_{51} = - 4 g_{31} - 3 g_{41}/2,   {\rm \hspace{.2in}}
g_{61} = 2 t z_t/T D,   {\rm \hspace{.2in}}   g_{71} = 0;   \\
\nonumber
g_{12} &=& 4 \beta (1+(m^2 z_2/s-u)/s\beta^2 - 2 t^2 u/s D)/D,
{\rm \hspace{.2in}}
g_{22} = (g_{32} + g_{42}/2)/2,    \\
\nonumber
g_{32} &=& 2 (-2 m^2 t (6 D-t z_2)/s^2\beta + (2 m^2 s+t^2)
             \beta + 2 m^2 t^2 z_2 \beta/D )/D^2,         \\
\nonumber
g_{42} &=& 4 (2 m^2 (m^2-s) + z_t^2 - 2 m^2 s t^2 \beta^2/D)/D^2,
{\rm \hspace{.2in}}   g_{52} = - 4 g_{32} - 3 g_{42}/2,    \\
\nonumber
g_{62} &=& 4 (4 m^2 s - t u (3+4 t/s) + 3 m^2 t^2/T)/D^2,
{\rm \hspace{.2in}}
g_{72} = 4 u/s D;                                      \\
\nonumber
g_{13} &=& -4 (D (m^2+6 m^2 t/s+s)/s\beta + 2 t U \beta)/D^2,
{\rm \hspace{.2in}}
g_{23} = (g_{33} + g_{43}/2)/2,    \\
\nonumber
g_{33} &=& -2 ( (8 m^2+t) D/s\beta + 2 m^2 t^2 z_2/s^2\beta 
               - t (3 m^2+t^2/s)\beta) + 2 m^2 t u z_2 \beta/D )/D^2,  \\
\nonumber
g_{43} &=& 4 (6 m^4-t z_t +2 m^2 s t u \beta^2/D)/D^2,  {\rm \hspace{.2in}}
g_{53} = - 4 g_{33} - 3 g_{43}/2,    \\
\nonumber
g_{63} &=& 4 (4 D+2 t (2+t/s) z_2 - 3 t^2 z_2/T + t^3 z_t/T^2)/D^2,
{\rm \hspace{.2in}}     g_{73} = -4 (1 - m^2 u/D)/s T;             \\
\nonumber
g_{14} &=& 4 (m^2 D (2 t/s-1)/\beta + 2 t^3 \beta)/s D^2,
{\rm \hspace{.2in}}           g_{24} = (g_{34} + g_{44}/2)/2,    \\
\nonumber
g_{34} &=& 2 ( 2 (2 m^2 u/s+z_2) D/s\beta - 2 m^2 t^2 z_2/s^2\beta
           + t z_t \beta + 2 t^3 z_t \beta/D )/D^2,        \\
\nonumber
g_{44} &=& 4 (2 T (3 m^2-s)+t^2 -2 s t^2 T \beta^2/D)/D^2,
{\rm \hspace{.2in}}   g_{54} = - 4 g_{34} - 3 g_{44}/2,    \\
\nonumber
g_{64} &=& 8 (s T - 2 t^2 u/s)/D^2,   {\rm \hspace{.2in}}
g_{74} = - 4 t/s D;                                          \\
\nonumber
g_{15} &=& 4 (D (m^2 z_2/s-t)/\beta - 2 t^2 u \beta)/s D^2,
{\rm \hspace{.2in}}           g_{25} = (g_{35} + g_{45}/2)/2,    \\
\nonumber
g_{35} &=& 2 ( 2 u (2 m^2-s) D/s^2\beta + t (2 T-t u /s)
            \beta + t^2 z_t/s\beta + 2 m^2 t^2 z_2 \beta/D )/D^2,  \\
\nonumber
g_{45} &=& 4 (6 m^4-2 m^2 s-t^2 + 2 s t T u \beta^2/D)/D^2,
{\rm \hspace{.2in}}   g_{55} = - 4 g_{35} - 3 g_{45}/2,    \\
\nonumber
g_{65} &=& 4 (2 D-t (1+4 t/s) u+m^2 t^2/T)/D^2,   {\rm \hspace{.2in}}
g_{75} = 4 u/s D.
\end{eqnarray}

%%%%%%%%%%%%%%%%%%%%%%%%%%%%%%%%%%%%%%%%%%%%%%%%%%%%%%%%%%%%%%%%

Now we list coefficients for the nonabelian box diagram (2a2):
\begin{eqnarray}
\nonumber
b_1 &=& -2/\varepsilon t,   {\rm \hspace{.2in}}
b_2 = 0,  {\rm \hspace{.2in}}
b_3 = (s/2 - t)/D,                   {\rm \hspace{.2in}}
b_4 = - (z_2/2 + z_t)/\beta D,       {\rm \hspace{.2in}}         \\  
\nonumber
b_5 &=& -10/t, {\rm \hspace{.2in}}  b_6=-2 z_t/t T, {\rm \hspace{.2in}}
b_7=2 (1/\varepsilon^2 + 1/\varepsilon + 2)/t,      {\rm \hspace{.2in}}
b_8 = -4/t,  {\rm \hspace{.2in}}   b_9 = 4/t;              \\
\nonumber
b_{11}^{(p)} &=& - 1/\varepsilon s - 2 (m^2 + t^2/s)/D,
{\rm \hspace{.2in}}
b_{21}^{(p)} = 1/2 s,   {\rm \hspace{.2in}}
b_{31}^{(p)} = t^2 (z_2/2 D - 2/s)/D,                      ß\\
\nonumber
b_{41}^{(p)} &=& (m^2 (s-4 t) D/s + s t z_t/2 - t^3 z_2/s)/\beta D^2,  
{\rm \hspace{.2in}}
b_{51}^{(p)} = -4/s,                              \\
\nonumber
b_{61}^{(p)} &=& - 8/\varepsilon s - 2 t (2 - t/T)/D, {\rm \hspace{.2in}}
b_{71}^{(p)} = (5/\varepsilon^2 + 2/\varepsilon + 4)/s,
{\rm \hspace{.2in}}  b_{81}^{(p)} = 8/s;                   \\
\nonumber
b_{12}^{(p)} &=& -4 m^2 (2 t/D - 1/s\beta^2)/D, {\rm \hspace{.2in}}
b_{22}^{(p)} = 0,  {\rm \hspace{.2in}}
b_{32}^{(p)} = 2 z_1 m^2 t/D^3,                    \\
\nonumber
b_{42}^{(p)} &=& -2 m^2 (2 m^2/s^2\beta^2 + 
     t (m^2 (s+4 t)+t^2)/D^2)/\beta D,   {\rm \hspace{.2in}}
b_{52}^{(p)} = 0,        \\
\nonumber
b_{62}^{(p)} &=& 4 ( m^2 (s+3 t) + m^4 t/T )/D^2,  {\rm \hspace{.2in}}
b_{72}^{(p)} = -4 u/s D,    {\rm \hspace{.2in}}  b_{82}^{(p)} = 0;  \\
\nonumber
b_{13}^{(p)} &=& 4 (2 m^2 u/D + (m^2+(3+4 t/s)(2 m^2-s))/s\beta^2)/D,
{\rm \hspace{.2in}}           b_{23}^{(p)} = 0,     \\
\nonumber
b_{33}^{(p)} &=& -z_t (2 m^2 s u/D-t)/D^2,         \\
\nonumber
b_{43}^{(p)} &=& ( 4 m^2 (m^2+z_2)/s^2\beta^2 - 2 m^2/s-2 m^2 s/D+
                             t^2/D - 2 m^4 s^2 \beta^2/D^2 )/\beta D,  \\
\nonumber
b_{53}^{(p)} &=& 0,  {\rm \hspace{.2in}}
b_{63}^{(p)} = 4 (D t^2/T^2 + 2 t^2 (s-t)/T + 
               10 m^2 s+4 t z_t-4 t^2 u/s)/D^2,                \\
\nonumber
b_{73}^{(p)} &=& 4 (2+t/s - t/T)/D,  {\rm \hspace{.2in}}
b_{83}^{(p)} = 0;                                             \\
\nonumber
b_{14}^{(p)} &=& 4 ((3 m^2+2 t)/\beta^2 + 2 t^3/D)/s D,
{\rm \hspace{.2in}}           b_{24}^{(p)} = 0,    {\rm \hspace{.2in}}
b_{34}^{(p)} = -t (2 m^2+3 t + 2 t^2 z_t/D)/D^2,     \\
\nonumber
b_{44}^{(p)} &=&  - ( 3 m^2+t+(3 m^2+2 t)/\beta^2 +
              t (m^2 s-3 t^2)/D + 2 s t^4 \beta^2/D^2 )/s\beta D,  \\
\nonumber
b_{54}^{(p)} &=& 0,                     {\rm \hspace{.2in}}
b_{64}^{(p)} = -8 t^2 z_2/s D^2,        {\rm \hspace{.2in}}
b_{74}^{(p)} = 4 t/s D,  {\rm \hspace{.2in}}  b_{84}^{(p)} = 0;   \\
\nonumber
b_{15}^{(p)} &=& -4 m^2 (2 t/D - 1/s\beta^2)/D,  {\rm \hspace{.2in}}
b_{25}^{(p)} = 0,    {\rm \hspace{.2in}}
b_{35}^{(p)} = 2 m^2 t z_1/D^3,                               \\
\nonumber
b_{45}^{(p)} &=& - 2 m^2 (2 m^2/s^2\beta^2 + 
           t (m^2 z_2+t z_t)/D^2)/\beta D,    {\rm \hspace{.2in}}
b_{55}^{(p)} = 0,                                        \\
\nonumber
b_{65}^{(p)} &=& 4 (m^2 s+4 m^2 t - t^2 + t^3/T)/D^2,  {\rm \hspace{.2in}}
b_{75}^{(p)} = -4 u/s D,  {\rm \hspace{.2in}}  b_{85}^{(p)} = 0;    \\
\nonumber
b_{11}^{(m)} &=& 2/\varepsilon s + (T-t^2/s-4 m^2 z_t/s\beta^2)/D,
{\rm \hspace{.2in}}
b_{21}^{(m)} = -1/s,                                           \\
\nonumber
b_{31}^{(m)} &=& t (z_2/s + t (2 m^2+t/2)/D)/D,           \\
\nonumber
b_{41}^{(m)} &=& ( 2 m^2 D/s\beta^2 - m^2 t z_2/s\beta^2 + 2 m^2 z_2+
          t (2 m^2-u) +3 s t^3 \beta^2/2 D+t^3 z_t/D )/s\beta D,  \\
\nonumber
b_{51}^{(m)} &=& 8/s,  {\rm \hspace{.2in}}
b_{61}^{(m)} = 8/\varepsilon s + 2 t (3 + 4 t/s)/D,   {\rm \hspace{.2in}}
b_{71}^{(m)} = -3 (2/\varepsilon^2+1/\varepsilon + 2 )/s,
{\rm \hspace{.2in}}     b_{81}^{(m)} = -8/s;                   \\
\nonumber
b_{12}^{(m)} &=& 2/\varepsilon s + (3 (m^2+t u/s) - 4 m^2 t z_2/s^2\beta^2)/D,
{\rm \hspace{.2in}}     b_{22}^{(m)} = -1/s,          \\
\nonumber
b_{32}^{(m)} &=& t (1+4 t/s + t u (3/2+2 t/s)/D)/D,   \\
\nonumber
b_{42}^{(m)} &=& -( T-t z_t/s - m^2 t z_2/s^2\beta^2 + t^2 u z_2/2 s D
                                      + m^2 s t \beta^2/D )/\beta D,  \\
\nonumber
b_{52}^{(m)} &=& 8/s,  {\rm \hspace{.2in}}
b_{62}^{(m)} = 2((1/t+4/s)/\varepsilon - 2(z_1/t-t z_2/s)/D + m^2 t/T D), \\
\nonumber
b_{72}^{(m)} &=& -(1/\varepsilon^2 t+6/\varepsilon^2 s - 2/\varepsilon t 
               + 3/\varepsilon s - 2 (2/t - 3/s)),   {\rm \hspace{.2in}}
b_{82}^{(m)} = - 2 (1/t + 4/s);                          \\
\nonumber
b_{11}^{(n)} &=& -2/\varepsilon s + (5 m^2 - 2 t + t^2/s + 4z_t/\beta^2 + 
               t z_2/s\beta^2)/D,        {\rm \hspace{.2in}}
b_{21}^{(n)} = 1/s,                       \\
\nonumber
b_{31}^{(n)} &=& - (2 m^2-3 t)/D + t^2 u (1/2-2 u/s)/D^2,   \\
\nonumber
b_{41}^{(n)} &=& -(4 z_t (s z_t - 3 m^2 t)/s^2\beta^2
       - 6 m^4 z_2/D - m^2 s t \beta^2/D - t (2 m^2/s+1))/2\beta D,  \\
\nonumber
b_{51}^{(n)} &=& - b_{52}^{(m)},        {\rm \hspace{.2in}}
b_{61}^{(n)} = - b_{62}^{(m)} - 16 m^4/T D,     {\rm \hspace{.2in}}
b_{71}^{(n)} = - b_{72}^{(m)},   {\rm \hspace{.2in}}
b_{81}^{(n)} = - b_{82}^{(m)};                           \\
\nonumber
b_{12}^{(n)} &=& -2/\varepsilon s -((8 m^2+s) z_t/s\beta^2 - 
                       m^2 - t^2/s)/D,    {\rm \hspace{.2in}}
b_{22}^{(n)} = 1/s,                         \\
\nonumber
b_{32}^{(n)} &=& -(2/s - s t T/D^2 - t^2 (2 m^2+t/2)/D^2),    \\
\nonumber
b_{42}^{(n)} &=& (3 m^2 z_t/s\beta^2 - 
                 t (2 T/s+1-(2 m^2-t/2) z_t/D))/\beta D,            \\
\nonumber
b_{52}^{(n)} &=& - b_{51}^{(m)},        {\rm \hspace{.2in}}
b_{62}^{(n)} = - b_{61}^{(m)} - 16 m^2/D,     {\rm \hspace{.2in}}      
b_{72}^{(n)} = - b_{71}^{(m)},   {\rm \hspace{.2in}}
b_{82}^{(n)} = - b_{81}^{(m)};                           \\
\nonumber
c_1 &=& -2/s\beta^2,    {\rm \hspace{.2in}}   c_2 = 0,  {\rm \hspace{.2in}}
c_3 = -s/2 D,   {\rm \hspace{.2in}}   c_4 = 1/2 s \beta^3 + z_2/2\beta D,  
{\rm \hspace{.2in}}   c_5 = 0,                              \\
\nonumber
c_6 &=& -2/T,      {\rm \hspace{.2in}}   c_7 = c_8 = 0;           \\
d_{11} &=& -8 z_t/s\beta^2 D,       {\rm \hspace{.2in}}
d_{21} = 0,  {\rm \hspace{.2in}}   d_{31} = (-s T + t^2)/D^2,      \\
\nonumber
d_{41} &=& (z_2/s\beta^2 + 2 t z_t/D)/\beta D,      {\rm \hspace{.2in}}
d_{51} = 0,   {\rm \hspace{.2in}}   d_{61} = -8/D,   {\rm \hspace{.2in}}
d_{71} = d_{81} = 0;                      \\
\nonumber
d_{12} &=& -8 z_u/s\beta^2 D,       {\rm \hspace{.2in}}
d_{22} = 0,  {\rm \hspace{.2in}}   d_{32} = -(m^2 s + t u)/D^2,     \\
\nonumber
d_{42} &=& z_2 (-1/s\beta^2 + 2 m^2/D)/\beta D,      {\rm \hspace{.2in}}
d_{52} = 0,   {\rm \hspace{.2in}}   d_{62} = -8 m^2/T D,
{\rm \hspace{.2in}}        d_{72} = d_{82} = 0;                      \\
\nonumber
g_{11} &=& 2 z_t/\beta^2 D,    {\rm \hspace{.2in}}
g_{21} = 0,   {\rm \hspace{.2in}}   g_{31} = - s t^2/2 D^2,       \\
\nonumber
g_{41} &=& -( 2/s+m^2 z_2/s\beta^2 D + s t z_t/2 D^2 )/\beta,
{\rm \hspace{.2in}}        g_{51} = 0,     {\rm \hspace{.2in}}
g_{61} = -2 t^2/T D,   {\rm \hspace{.2in}}   g_{71} = g_{81} = 0;   \\
\nonumber
g_{12} &=& -4( 2+t/s-2 U/s\beta^2-3 m^2 z_2/s^2\beta^4-2 m^2 t 
                z_2/s\beta^2 D )/D,        {\rm \hspace{.2in}}
g_{22} = 0,                          \\
\nonumber
g_{32} &=& s( 4 m^2+t z_t/s + 2 t^3 u/s D )/D^2,     \\
\nonumber
g_{42} &=& ( 2 (2 m^4+2 m^2 u-t T)/D +
                2 (4 m^4/s+m^2+2 m^2 t/s-t)/s\beta^2+
                 t^2 z_2/s\beta^2 D                       \\
\nonumber
&&  {\rm \hspace{3.in}} 
        - 12 m^4 z_2/s^3\beta^4 - 2 t^3 z_u/D^2 )/\beta D,  \\
\nonumber
g_{52} &=& 0,  {\rm \hspace{.2in}}
g_{62} = 4( 2 D(3+t/s)+3 t u + 5 m^2 t^2/T )/D^2,   {\rm \hspace{.2in}}
g_{72} = 4 z_u/s\beta^2 D,     {\rm \hspace{.2in}}  g_{82} = 0;  \\
\nonumber
g_{13} &=& -4( 2+t/s+(2 m^2-u)/s\beta^2+3 m^2 z_2/s^2\beta^4-2 t U 
               z_2/s\beta^2 D )/D,         {\rm \hspace{.2in}}
g_{23} = 0,                          \\
\nonumber
g_{33} &=& s( 4 m^2+t z_t/s - 2 m^2 t u/D)/D^2,               \\
\nonumber
g_{43} &=& ( 2 (2 m^4-s z_t-2 t^2+t^3/s)/D +
                2 (4 m^4/s+3 m^2+16 m^2 t/s)/s\beta^2-
                 t^2 z_2/s\beta^2 D                       \\
\nonumber
&&  {\rm \hspace{2.5in}}
  + 12 m^4 z_2/s^3\beta^4 - 2 t^2 (t z_u+z_2 u)/D^2 )/\beta D,   \\
\nonumber
g_{53} &=& 0,  {\rm \hspace{.2in}}
g_{63} = 4 (2 (2+t/s) + 2 t^2 U/T D - t^2/T^2)/D,   {\rm \hspace{.2in}}
g_{73} = 4 (t/T + z_t/s\beta^2)/D,                     \\
\nonumber
g_{83} &=& 0;                                        \\
\nonumber
g_{14} &=& 4( 2 m^2 (s-2 u)/s^2\beta^2 - 3 m^2 z_2/s^2\beta^4 + 
             2 t^2 z_t/s\beta^2 D )/D,         {\rm \hspace{.2in}}
g_{24} = 0,                          \\
\nonumber
g_{34} &=& -(2 T u+t^2 + 2 t^4/D)/D^2,       \\
\nonumber
g_{44} &=& ( 3 t^2/D + (3 m^2+18 m^2 t/s-2 s-8 t)/s\beta^2-
               8 m^4 z_1/s^2\beta^2 D +3 m^2 z_2/s^2\beta^4     \\
\nonumber
&&  {\rm \hspace{4.in}}
        + 2 t^4 z_2/s D^2 )/\beta D,                     \\
\nonumber
g_{54} &=& 0,  {\rm \hspace{.2in}}
g_{64} = 8( 2+t/s + t^2/D )/D,     {\rm \hspace{.2in}}
g_{74} = 4 z_t/s\beta^2 D,   {\rm \hspace{.2in}}   g_{84} = 0;     \\
\nonumber
g_{15} &=& 4( 2 m^2 z_2/s+3 T/\beta^2-6 m^2 t/s\beta^2 + 2 t^2 z_u/D 
                        )/s\beta^2 D,     {\rm \hspace{.2in}}
g_{25} = 0,                          \\
\nonumber
g_{35} &=& 2/D + t z_t/D^2 + 2 t^3 u/D^3,     \\
\nonumber
g_{45} &=& ( 2 (2 m^4+t^2 u/s-m^2 t^2 z_2/D)/D + t^2 z_2/s\beta^2 D-
           2 (2 m^2 (m^2-s)+(m^2+s) z_2                  \\
\nonumber
&&  {\rm \hspace{3.5in}}
              + 12 m^4 z_t/s\beta^2)/s^2\beta^2 )/\beta D,   \\
\nonumber
g_{55} &=& 0,  {\rm \hspace{.2in}}
g_{65} = 4( -2 D u/s + t z_2 - t^3/T )/D^2,    {\rm \hspace{.2in}}
g_{75} = 4 z_u/s\beta^2 D,   {\rm \hspace{.2in}}   g_{85} = 0;
\end{eqnarray}

%%%%%%%%%%%%%%%%%%%%%%%%%%%%%%%%%%%%%%%%%%%%%%%%%%%%%%%%%%%%%%%%

Finally, the coefficients for the crossed box (2a4) are:
\begin{eqnarray}
\nonumber
b_1 &=& 2 (2-u/t)/s,   {\rm \hspace{.2in}}  b_2 = b_1(t\leftrightarrow u),
{\rm \hspace{.2in}}  b_3 = (2 m^2-3 t u/s+s)/D,  {\rm \hspace{.2in}}
b_4 = 0,                          \\
\nonumber
b_5 &=& 8 (3/s - s/t u),  {\rm \hspace{.2in}}
b_6 = 2/\varepsilon u + z_t/t T,   {\rm \hspace{.2in}}
b_7 = b_6(t\leftrightarrow u),   {\rm \hspace{.2in}}
b_8 = -b_5/2,                       \\
\nonumber
b_9 &=& (1/\varepsilon^2 + 1/\varepsilon + 2) s/t u;      \\
\nonumber
b_{11}^{(p)} &=& - 2 (z_u/u + 2)/s,   {\rm \hspace{.2in}}
b_{21}^{(p)} = -2 (z_u - 2 t)/s u,   {\rm \hspace{.2in}}
b_{31}^{(p)} = - 2 z_u/s u + 2 t/D - t u^2 z_t/s D^2,     \\
\nonumber
b_{41}^{(p)} &=& -4 m^2/t u,     {\rm \hspace{.2in}}
b_{51}^{(p)} = 4 b_{21}^{(p)},    {\rm \hspace{.2in}}
b_{61}^{(p)} = - 4/\varepsilon u - 2 m^2 z_t/T D,        \\
\nonumber
b_{71}^{(p)} &=& 2 z_u (t/u+2 m^2/U)/D,      {\rm \hspace{.2in}}
b_{81}^{(p)} = -2 b_{21}^{(p)},    {\rm \hspace{.2in}}
b_{91}^{(p)} = 2 (1/\varepsilon^2 + 1/\varepsilon + 2)/u;     \\
\nonumber
b_{12}^{(p)} &=& 8 m^2/s^2 u,     {\rm \hspace{.2in}}
b_{22}^{(p)} = b_{12}^{(p)},     {\rm \hspace{.2in}}
b_{32}^{(p)} = 4 m^2 (2 m^2 s (D/u-t) D + t u^2 z_1)/s^2 D^3,   \\
\nonumber
b_{42}^{(p)} &=& -8 m^2/t^2 u,     {\rm \hspace{.2in}}
b_{52}^{(p)} = 4 b_{12}^{(p)},     {\rm \hspace{.2in}}
b_{62}^{(p)} = -4 m^2 ( 2 (1/t-1/s) D-u (3 t-u)/s + m^2 t/T )/D^2,  \\
\nonumber
b_{72}^{(p)} &=& -4 m^2 (m^4/U+m^2-t^2/s+t u/s)/D^2,  {\rm \hspace{.2in}}
b_{82}^{(p)} = -2 b_{12}^{(p)},      {\rm \hspace{.2in}}
b_{92}^{(p)} = 4 z_t/t D;                                \\
\nonumber
b_{13}^{(p)} &=& 4 (2 m^2/u - 3)/s^2,       {\rm \hspace{.2in}}
b_{23}^{(p)} = b_{13}^{(p)},                       \\
\nonumber
b_{33}^{(p)} &=& b_{13}^{(p)} - 2 ( 2 t u/D+(4 m^2 u^3-t u^2 z_t)/D^2
                            - 2 m^2 t u^3 (t - u)/D^3)/s^2,    \\
\nonumber
b_{43}^{(p)} &=& -8 m^2/t^2 u,     {\rm \hspace{.2in}}
b_{53}^{(p)} = 4 b_{13}^{(p)},                                 \\
\nonumber
b_{63}^{(p)} &=&-4 (2 m^2 ((1/t-1/s) D+u (t+5 u)/s) + t u (5 m^4+3 t u)/sT + 
                            m^2 t^2 (m^4+t u)/s T^2)/D^2,     \\
\nonumber
b_{73}^{(p)} &=& -4 (2 m^2+5 u+2 t u z_u/D)/s D,    {\rm \hspace{.2in}}
b_{83}^{(p)} = -2 b_{13}^{(p)},        {\rm \hspace{.2in}}
b_{93}^{(p)} = 4 m^2 z_t/t T D;                                \\
\nonumber
b_{14}^{(p)} &=& 4 (2 m^2 t/u^2 + 4 m^2/u + 3)/s^2,    {\rm \hspace{.2in}}
b_{24}^{(p)} = b_{14}^{(p)},                          \\
\nonumber
b_{34}^{(p)} &=& b_{14}^{(p)} + 2 (t (4 m^2 t (t-u)-u^2 (2 m^2+3 t)) D
                       + 2 m^2 t^3 u (t-u))/s^2 D^3,             \\
\nonumber
b_{44}^{(p)} &=& -8 m^2/t u^2,          {\rm \hspace{.2in}}
b_{54}^{(p)} = 4 b_{14}^{(p)},          {\rm \hspace{.2in}}
b_{64}^{(p)} = 4 (2 m^2+5 t+2 t u z_t/D)/s D,             \\
\nonumber
b_{74}^{(p)} &=& 4 (2 m^2 ((s/u-1) D+t (u+5 t)) + u t (5 m^4+3 t u)/U + 
                             m^2 u^2 (m^4+t u)/U^2)/s D^2,     \\
\nonumber
b_{84}^{(p)} &=& -2 b_{14}^{(p)},      {\rm \hspace{.2in}}
b_{94}^{(p)} = -4 m^2 z_u/u U D;                                \\
\nonumber
b_{15}^{(p)} &=& 8 m^2 (t + 2 u)/s^2 u^2,       {\rm \hspace{.2in}}
b_{25}^{(p)} = b_{15}^{(p)},                            \\
\nonumber
b_{35}^{(p)} &=& b_{15}^{(p)} - 4 m^2 (t u (t-2 u) D + 
                                            t^2 u^2 (t-u))/s^2 D^3,
{\rm \hspace{.2in}}     b_{45}^{(p)} = -8 m^2/t u^2,         \\
\nonumber
b_{55}^{(p)} &=& 4 b_{15}^{(p)},          {\rm \hspace{.2in}}
b_{65}^{(p)} = 4 m^2 (2 m^2 s-u^2 + t^2 U/T + 2 m^2 t u/T)/s D^2,   \\
\nonumber
b_{75}^{(p)} &=& 4 m^2 (-2 m^2 s (t/u+2)+3 t^2 - u^2 (m^2-t)/U)/s D^2,
{\rm \hspace{.2in}}    b_{85}^{(p)} = -2 b_{15}^{(p)},
{\rm \hspace{.2in}}    b_{95}^{(p)} = -4 z_u/u D;             \\
\nonumber
b_{11}^{(m)} &=& -2 (2 m^2 + 3 t u/s)/s u,   {\rm \hspace{.2in}}
b_{21}^{(m)} = b_{11}^{(m)} + 2/u,                       \\
\nonumber
b_{31}^{(m)} &=& (-2 (2 m^2 s/u+t) + 2 s (2 m^2 u-s t)/D - t^3 u^2/D^2)/s^2,
{\rm \hspace{.2in}}        b_{41}^{(m)} = -4 m^2/t u,       \\
\nonumber
b_{51}^{(m)} &=& 4 b_{11}^{(m)} + 16/u,       {\rm \hspace{.2in}}
b_{61}^{(m)} = 4/\varepsilon u - 2 (2 D - t u)/s D,       \\
\nonumber
b_{71}^{(m)} &=& 2/\varepsilon u + 2 (2 D + t^2)/s D,
{\rm \hspace{.2in}}   b_{81}^{(m)} = -b_{51}^{(m)}/2,  {\rm \hspace{.2in}}
b_{91}^{(m)} = -3/\varepsilon^2 u;                       \\
\nonumber
b_{12}^{(m)} &=& 2 (-2 m^2 s/u-5 t-u+u^2/t)/s^2,   {\rm \hspace{.2in}}
b_{22}^{(m)} = 2 (-2 m^2/s - 1 + 3 t^2/s^2)/u,                      \\
\nonumber
b_{32}^{(m)} &=& -(2 s (2 m^2/u-1) + 2 t (3 m^2 s+t^2)/D - t^2 u^3/D^2)/s^2,
{\rm \hspace{.2in}}     b_{42}^{(m)} = b_{41}^{(m)},         \\
\nonumber
b_{52}^{(m)} &=& 4 b_{22}^{(m)},     {\rm \hspace{.2in}}
b_{62}^{(m)} = 2 (2/u-1/t)/\varepsilon - 2 (2 m^2 (2+u/t)+t (t-2 u)/s 
                      + t^2/T)/D,                           \\
\nonumber
b_{72}^{(m)} &=& 2 (t (-2 m^2 s/u+t-2 u)/s + m^2 u/U)/D,
{\rm \hspace{.2in}}   b_{82}^{(m)} = -2 b_{22}^{(m)},          \\
\nonumber
b_{92}^{(m)}&=&(1/t-2/u)/\varepsilon^2 + 2 s/\varepsilon t u + 4 s/t u; 
\\
b_{11}^{(n)} &=& 6 (2 t + u)/s^2,          {\rm \hspace{.2in}}
b_{21}^{(n)} = -2 (3 t^2/s^2 - 1)/u,                        \\
\nonumber
b_{31}^{(n)} &=& (2 (2 t+3 u) + 2 t (t^2+t u+5 u^2)/D + 5 t^2 u^3/D^2)/s^2,
{\rm \hspace{.2in}}    b_{41}^{(n)} = 0,          \\
\nonumber
b_{51}^{(n)} &=& 4 b_{21}^{(n)},    {\rm \hspace{.2in}}
b_{61}^{(n)} = -4/\varepsilon u + 2 (D/s + 2 u^2/s + 7 m^4/T)/D,
\\             \nonumber
b_{71}^{(n)} &=& 2 (t (2 - t/u) z_u+6 m^2 u - m^2 s u/U)/s D,
{\rm \hspace{.2in}}    b_{81}^{(n)} = -2 b_{21}^{(n)},  {\rm \hspace{.2in}}
b_{91}^{(n)} = 2 (1/\varepsilon^2 + 1/\varepsilon + 2)/u;      \\
\nonumber
b_{12}^{(n)} &=& 6 t/s^2,  {\rm \hspace{.2in}}
b_{22}^{(n)} = -2 (2/u - 3 t/s^2),                 {\rm \hspace{.2in}}
b_{32}^{(n)} = t (2 (t^2+u^2)/D - 2 - t u (6 m^2 s-t u)/D^2)/s^2,  \\
\nonumber
b_{42}^{(n)} &=& 0,    {\rm \hspace{.2in}}
b_{52}^{(n)} = 4 b_{22}^{(n)},   {\rm \hspace{.2in}}
b_{62}^{(n)} = -4/\varepsilon u + 2 (8 m^2 - 3 t u/s)/D,     \\
\nonumber
b_{72}^{(n)} &=& -2 (t (2 m^2 t/u-4 m^2 + 3 t) - 6 m^2 u (m^2-t)/U)/s D,
{\rm \hspace{.2in}}   b_{82}^{(n)} = -2 b_{22}^{(n)},  {\rm \hspace{.2in}}
b_{92}^{(n)} = b_{91}^{(n)};                \\
\nonumber
c_1 &=& c_2 = 0,   {\rm \hspace{.2in}}   c_3 = (u-t)/D,  {\rm \hspace{.2in}}
c_4 = c_5 = 0,     {\rm \hspace{.2in}}   c_6 = 1/T,    {\rm \hspace{.2in}}
c_7 = -1/U,    {\rm \hspace{.2in}}    c_8 = c_9 = 0;       \\
\nonumber
d_{11} &=& d_{21} = 0,  {\rm \hspace{.2in}}  d_{31} = 2 T u/D^2,
{\rm \hspace{.2in}}    d_{41} = d_{51} = 0,   {\rm \hspace{.2in}}
d_{61} = 4/D,    {\rm \hspace{.2in}}   d_{71} = -4 m^2/U D,    \\
\nonumber
d_{81} &=& d_{91} = 0;                   \\
\nonumber
d_{12} &=& d_{22} = 0,  {\rm \hspace{.2in}}  d_{32} = -2 t U/D^2,
{\rm \hspace{.2in}}    d_{42} = d_{52} = 0,   {\rm \hspace{.2in}}
d_{62} = 4 m^2/T D,   {\rm \hspace{.2in}}   d_{72} = -4/D,     \\
\nonumber
d_{82} &=& d_{92} = 0;                \\
\nonumber
g_{11} &=& 6/s,   {\rm \hspace{.2in}}  g_{21} = g_{11}, {\rm \hspace{.2in}}
g_{31} = (6 - 2 t^2/D - t^2 u^2/D^2)/s,   {\rm \hspace{.2in}}
g_{41} = 0,    {\rm \hspace{.2in}}   g_{51} = 4 g_{11},        \\
\nonumber
g_{61} &=& -2 m^2 t/T D,    {\rm \hspace{.2in}}
g_{71} = -2 (t + 2 m^2 u/U)/D,     {\rm \hspace{.2in}}
g_{81} = -2 g_{11},    {\rm \hspace{.2in}}    g_{91} = 0;      \\
\nonumber
g_{12} &=& -12/s^2, {\rm \hspace{.2in}}  g_{22} = g_{12}, {\rm \hspace{.2in}}
g_{32} = -2 (6 - u^2 (4 m^2 s+5 t^2)/D^2 - 2 t^3 u^3/D^3)/s^2,    \\
\nonumber
g_{42} &=& 0,   {\rm \hspace{.2in}}   g_{52} = 4 g_{12}, {\rm \hspace{.2in}}
g_{62} = 4 (5 m^4 t/T + 3 u (2 m^2-t u/s))/D^2,                  \\
\nonumber
g_{72} &=& -4 (m^2 (2 t+3 u)-t u (t+4 u)/s + m^2 u^2/U)/D^2,
{\rm \hspace{.2in}}     g_{82} = -2 g_{12},   {\rm \hspace{.2in}}
g_{92} = 4/D;                                                 \\
\nonumber
g_{13} &=& g_{12},  {\rm \hspace{.2in}}  g_{23} = g_{12}, {\rm \hspace{.2in}}
g_{33} = g_{32} + 4 m^4 s u/D^3,      {\rm \hspace{.2in}}
g_{43} = 0,    {\rm \hspace{.2in}}   g_{53} = g_{52},           \\
\nonumber
g_{63} &=& 4 (u (4 D + t u)/s + 3 m^2 t U/T + m^4 t^2/T^2)/D^2,
{\rm \hspace{.2in}}
g_{73} = 4 u (2 t u - D)/s D^2,                              \\
\nonumber
g_{83} &=& -2 g_{12},      {\rm \hspace{.2in}}   g_{93} = 4 m^2/T D;    \\
\nonumber
g_{14} &=& g_{12},  {\rm \hspace{.2in}}  g_{24} = g_{12},  {\rm \hspace{.2in}}
g_{34} = g_{33} + 4 t^2 u z_u/D^3,    {\rm \hspace{.2in}}  g_{44} = 0,
{\rm \hspace{.2in}}         g_{54} = g_{52},                           \\
\nonumber
g_{64} &=& 4 (3 t+4 u + 2 t^2 u/D)/s D,     {\rm \hspace{.2in}}
g_{74} = 4 u (-3 t^2/s + 3 m^2 t/U - m^6/U^2)/D^2,               \\
\nonumber
g_{84} &=& -2 g_{12},   {\rm \hspace{.2in}}
g_{94} = g_{93} (t\leftrightarrow u);           \\
\nonumber
g_{15} &=& g_{12}, {\rm \hspace{.2in}}  g_{25} = g_{12},  {\rm \hspace{.2in}}
g_{35} = g_{34} - 4 m^4 s t/D^3,    {\rm \hspace{.2in}}
g_{45} = 0,    {\rm \hspace{.2in}}   g_{55} = g_{52},         \\
\nonumber
g_{65} &=& 4 (m^4 t/T + u (2 m^2-3 t u/s))/D^2,    {\rm \hspace{.2in}}
g_{75} = g_{65} (t\leftrightarrow u),     {\rm \hspace{.2in}}
g_{85} = -2 g_{12},     {\rm \hspace{.2in}}    g_{95} = g_{92};
\end{eqnarray}

\vglue 1cm
\begin{center}\begin{large}\begin{bf}
APPENDIX B
\end{bf}\end{large}\end{center}
\vglue .3cm

\setcounter{equation}{0}
\renewcommand{\theequation}{B\arabic{equation}}

This Appendix contains the coefficients for the one-loop corrections to the 
subprocess $q \bar{q} \rightarrow Q \overline Q$. For the coefficients $h$ 
in (\ref{boxq}) for the box diagram (5a) we have:
\begin{eqnarray}
\nonumber
h_1 &=& (t^3/D + m^2 + 2 t^2/s)/2 D,   {\rm \hspace{.14in}}
h_2 = 2 (2/\varepsilon s + 2/s - t/D),   {\rm \hspace{.14in}}
h_3 =-2 (1 + t z_t/\beta^2 D)/s,                               \\
\nonumber
h_4 &=& - ( (m^2 s T + t^4/s)/D - t (2 T + z_t/\beta^2)/s )/2\beta D,
{\rm \hspace{.2in}}
h_5 = -4/s,   {\rm \hspace{.2in}}     h_6 = - 2/\varepsilon^2 s;   \\
%\nonumber
%\quad\\
\nonumber
h_1^{(1)} &=& 2 t T/D^2,   {\rm \hspace{.2in}}   h_2^{(1)} = 8 t/s D,
{\rm \hspace{.2in}}
h_3^{(1)} = - 8 (1 - t z_t/D)/s^2\beta^2,                         \\
h_4^{(1)} &=& 2 z_t ( T/D + 2 m^2/s^2\beta^2 )/\beta D;             \\
\nonumber
h_1^{(2)} &=& - z_1/4 D^2,    {\rm \hspace{.2in}}  h_2^{(2)} = - z_t/T D, 
{\rm \hspace{.2in}}    h_3^{(2)} = 1/D,   {\rm \hspace{.2in}}
h_4^{(2)} =(1 - s t \beta^2/D)/4\beta D;                           \\
\nonumber
h_1^{(3)} &=& t/8 D,    {\rm \hspace{.2in}}    h_2^{(3)} = 0,
{\rm \hspace{.2in}}   h_3^{(3)} = 0,   {\rm \hspace{.2in}} 
h_4^{(3)} = z_t/8\beta D;                                          \\
\nonumber
h_1^{(4)} &=& t^2/D^2,   {\rm \hspace{.2in}}   h_2^{(4)} = - 4/D,
{\rm \hspace{.2in}}    h_3^{(4)} = - 4 z_t/s\beta^2 D,  {\rm \hspace{.2in}}
h_4^{(4)} = ( t z_1/D + z_t/\beta^2 )/s\beta D;                    \\
\nonumber
h_1^{(5)} &=& s t/2 D^2,   {\rm \hspace{.2in}}    h_2^{(5)} = 2 t/T D, 
{\rm \hspace{.2in}}     h_3^{(5)} = 2 z_2/s\beta^2 D,  {\rm \hspace{.2in}}
h_4^{(5)} = - ( 2 z_t/s\beta^2 + t z_2/D )/2\beta D;               \\
\nonumber
h_1^{(6)} &=& h_1^{(5)}/2,   {\rm \hspace{.2in}}   h_2^{(6)} = h_2^{(5)}/2, 
{\rm \hspace{.2in}}    h_3^{(6)} = h_3^{(5)}/2,    {\rm \hspace{.2in}}
h_4^{(6)} = h_4^{(5)}/2;
\end{eqnarray}

The nontrivial coefficients for the second box diagram (5b) are:
\begin{eqnarray}
\nonumber
h_1 &=& (m^2 s t/D - m^2 - 4 u - 2 u^2/s)/2 D,  {\rm \hspace{.2in}}
h_2 = - 2 (2 (1/\varepsilon + 1) U/s + 1 - m^2 t/D)/U,        \\
\nonumber
h_3 &=& 2 (1 + z_t t/\beta^2 D)/s,                             \\
h_4 &=& ( s \beta^2 - 2 z_u - 2 D/s\beta^2 + 4 m^2 u z_u/s^2\beta^2
             + (m^2 z_{1u} + s u t \beta^2)/D )/2\beta D,       \\
\nonumber
h_5 &=& 4/s,  {\rm \hspace{.2in}}   h_6 = 2/\varepsilon^2 s;    \\
\nonumber
h_1^{(1)} &=& (m^2 s + u z_u)/D^2,    {\rm \hspace{.16in}} 
h_2^{(1)} = 4 (z_u/U + 2 u/s)/D,    {\rm \hspace{.16in}}
h_3^{(1)} = - 2 (1/\beta^2 + 1) (s+2u)/s D,                     \\
\nonumber
h_4^{(1)} &=& ((-m^2 s \beta^2 + u z_{1u}/s)/D
                        + (8 m^4/s + u)/s\beta^2)/\beta D,
\end{eqnarray}
where for convenience we have introduced the notation:
\begin{equation}  
z_{1u} \equiv m^2 s - u^2 = z_1 (t\rightarrow u).
\end{equation}

The remaining coefficients for this graph can be easily obtained from the 
corresponding ones for the graph (5a) through the relations (\ref{relsq}).

\newpage
%\vglue 1cm
\begin{center}\begin{large}\begin{bf}
REFERENCES
\end{bf}\end{large}\end{center}
\vglue .3cm

   \begin{list}{$[$\arabic{enumi}$]$} 
    {\usecounter{enumi} \setlength{\parsep}{0pt} 
     \setlength{\itemsep}{3pt} \settowidth{\labelwidth}{(99)} 
     \sloppy} 
\item \label{compass} 
G.~Baum et al, COMPASS Collaboration: CERN/SPLC 96-14 and 96-30.
\item \label{slac}
R. Arnold et al, SLAC-PROPOSAL-E156, 1997.
\item \label{desy}
A. de Roeck and T. Gehrmann, DESY-Proceedings-1998-1.
\item \label{Nason}
P.~Nason, S.~Dawson and R.~K.~Ellis, Nucl. Phys. {\bf B303}, 607 (1988); 
Nucl. Phys. {\bf B327}, 49 (1989); Erratum: {\it ibid} {\bf B335}, 260 
(1990). 
\item \label{Been}
W.~Beenakker, H.~Kuijf, W.~L.~van Neerven and J.~Smith, Phys. Rev. 
{\bf D40}, 54 (1989);
W.~Beenakker, W.~L.~van Neerven, R.~Meng, G.~A.~Schuler and J.~Smith, 
Nucl. Phys. {\bf B351}, 507 (1991).
\item \label{Ellis}
R.~K.~Ellis and P.~Nason, Nucl. Phys. {\bf B312}, 551 (1989).
\item \label{Smith}
J.~Smith and W.L.~van Neerven, Nucl. Phys. {\bf B374}, 36 (1992).
\item \label{Bojak1}
I.~Bojak and M.~Stratmann, hep-ph/0112276.
\item \label{Bojaka}
I.~Bojak and M.~Stratmanm, Phys. Lett. {\bf B433}, 411 (1998).
\item \label{Bojakb}
I.~Bojak and M.~Stratmanm, Nucl. Phys. {\bf B540}, 345 (1999);
Erratum: {\it ibid} {\bf B569}, 694 (2000).
\item \label{MCGa}
A.P.~Contogouris, Z.~Merebashvili and G.~Grispos, Phys. Lett. {\bf B482}, 
93 (2000).
\item \label{MCGb}
Z.~Merebashvili, A.P.~Contogouris and G.~Grispos, Phys. Rev. {\bf D62}, 
114509, (2000).
\item \label{Mirkes}
J.~H.~K\"{u}hn, E.~Mirkes and J.~Steegborn, Z. Phys. C {\bf 57}, 615, 
(1993). 
\item \label{Drees}
M.~Drees, M.~Kr\"{a}mer, J.~Zunft and P.~M.~Zerwas, 
Phys. Lett. {\bf B306}, 371 (1993).
\item \label{KMC}
B.~Kamal, Z.~Merebashvili and A.P.~Contogouris, Phys. Rev. {\bf D51}, 
4808 (1995); Erratum: {\it ibid} {\bf D55}, 3229 (1997).
\item \label{JT}
G.~Jikia and A.~Tkabladze, Phys. Rev. {\bf D54}, 2030 (1996).

\item \label{DREG}
G.'t~Hooft and M.~Veltman, Nucl. Phys. {\bf B44}, 189 (1972).
\item \label{DRED}
W.~Siegel, Phys. Lett. {\bf 84B}, 193 (1979).
\item \label{Slaven}
W.C.~Kuo, D.~Slaven and B.L.~Young, Phys. Rev. {\bf D37}, 233 (1988). 
\item \label{Muta} 
T.~Muta, "Foundations of Quantum Chromodynamics" 
(World Scientific, 1987).
\item \label{Andrei}
A.~I.~Davydychev, P.~Osland and O.~V.~Tarasov, Phys. Rev. {\bf D54},
4087 (1996); Erratum: {\it ibid} {\bf D59}, 109901 (1999).

\item \label{passar}
G.~Passarino and M.~Veltman, Nucl. Phys. {\bf B160}, 151 (1979).
\item \label{reduce}
A.~Hearn, REDUCE User's Manual Version 3.6
(Rand Corporation, Santa Monica, CA, 1995).

\item \label{Cutkosky}
R.~E.~Cutkosky, J. Math. Phys. {\bf 1}, 429 (1960).
\item \label{KMM}
J.~G.~K\"{o}rner, B.~Meli\'{c} and Z.~Merebashvili, Phys. Rev. {\bf D62}, 
096011 (2000).
\end{list}

%\newpage
\vglue 3cm
\begin{center}\begin{large}\begin{bf}
FIGURE CAPTIONS
\end{bf}\end{large}\end{center}
\vglue .3cm

\begin{list}{Fig.~\arabic{enumi}.} 
    {\usecounter{enumi} \setlength{\parsep}{0pt} 
     \setlength{\itemsep}{3pt} \settowidth{\labelwidth}{Fig.~9.} 
     \sloppy}
\item{}
The t-, u- and s-shannel leading order (Born) graphs contributing to the 
gluon (curly lines) fusion amplitude. The thick solid lines correspond to
the heavy quarks.
\item{}
The t-channel one-loop graphs contributing to the gluon fusion amplitude. 
Loops with dotted lines represent gluon, ghost and light and heavy quarks. 
\item{}
The s-channel one-loop graphs contributing to the gluon fusion amplitude. 
Loops with dotted lines represent gluon, ghost and light and heavy quarks.
\item{} 
The lowest order Feynman diagram contributing to the subprocess 
$q \bar{q} \rightarrow Q \overline Q$. The thick lines correspond 
to the heavy quarks. 
\item{}
The one-loop Feynman diagrams contributing to the subprocess 
$q \bar{q} \rightarrow Q \overline Q$.
The loop with dotted line represents gluon, ghost and light and heavy 
quarks.
\end{list}

\newpage
\pagestyle{empty}

\vglue 4cm
%\begin{figure}
\centering
\mbox{\epsffile{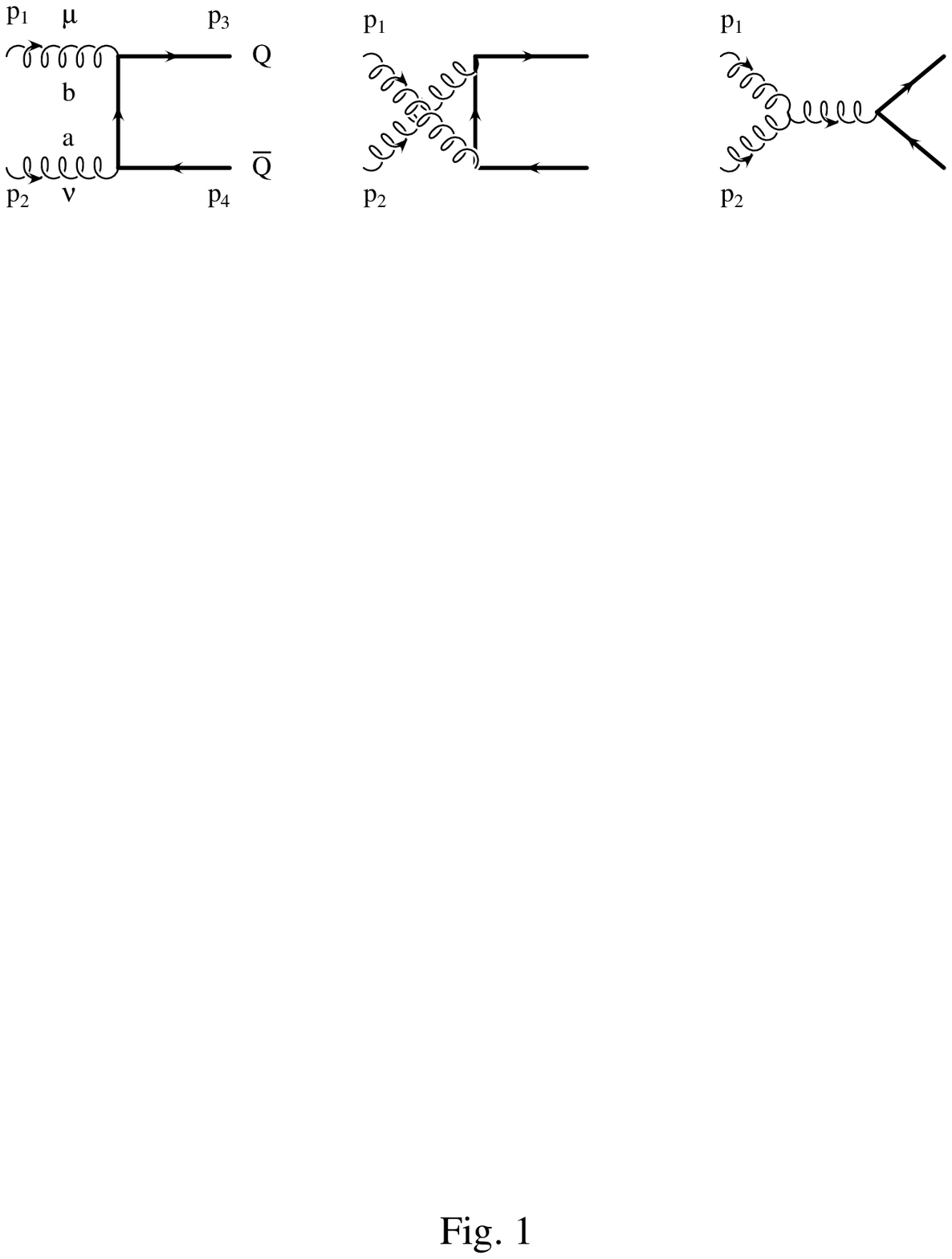}}
%\end{figure}

\newpage
%\begin{figure}
\centering 
\mbox{\epsffile{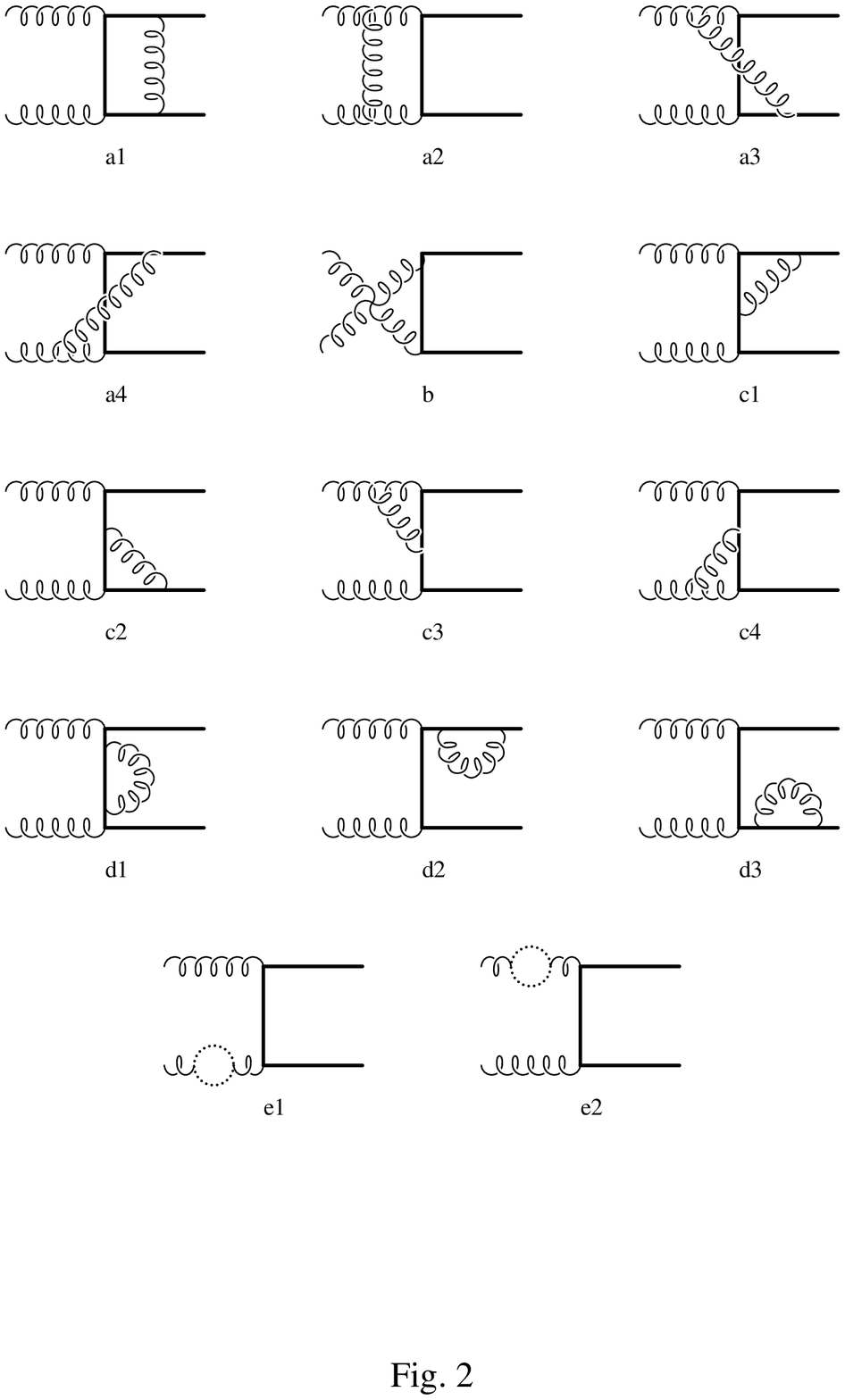}}
%\end{figure}

\newpage
%\begin{figure} 
\centering
\mbox{\epsffile{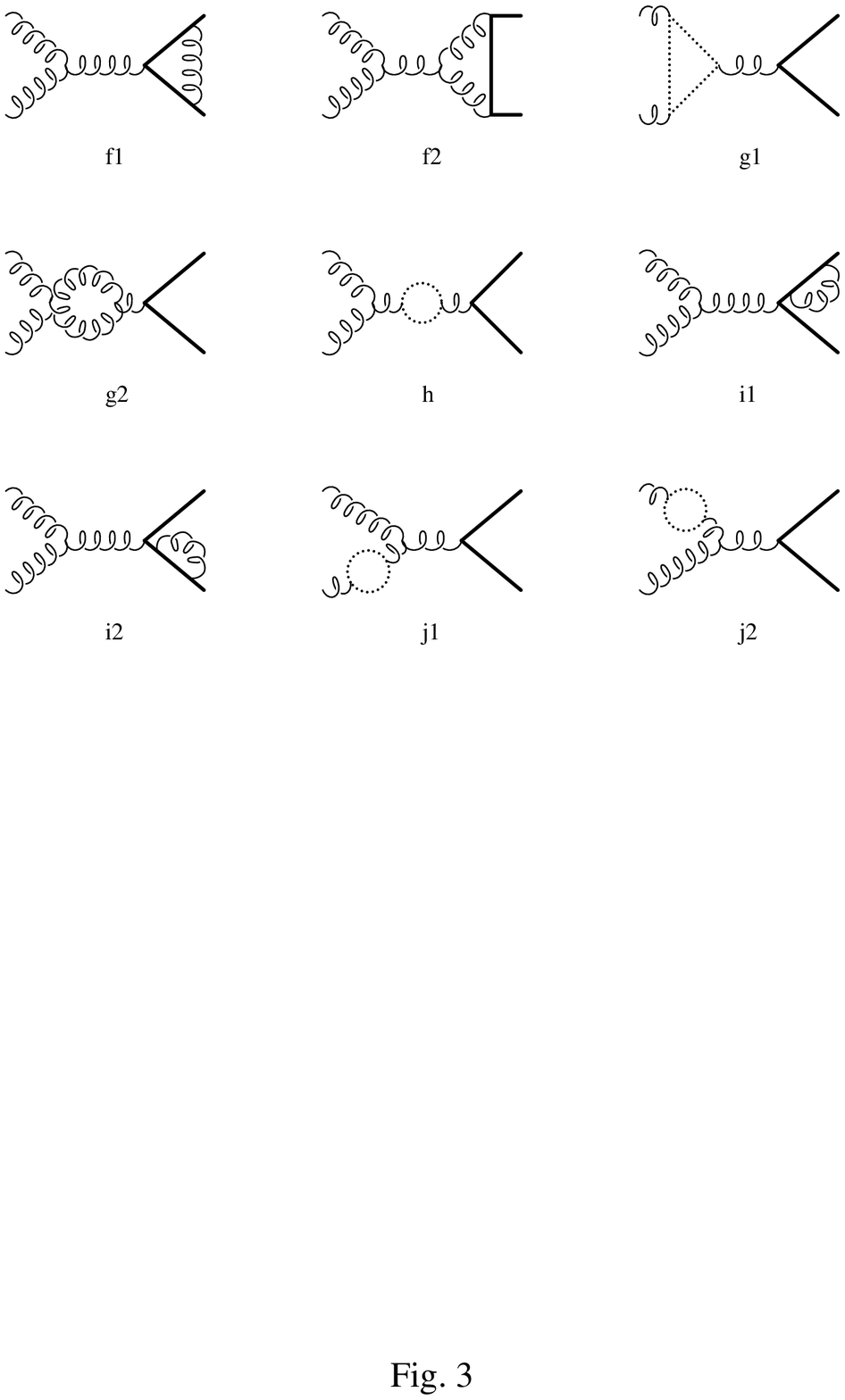}}
%\end{figure}

\newpage
\vglue 4.3cm
%\begin{figure} 
\centering
\mbox{\epsffile{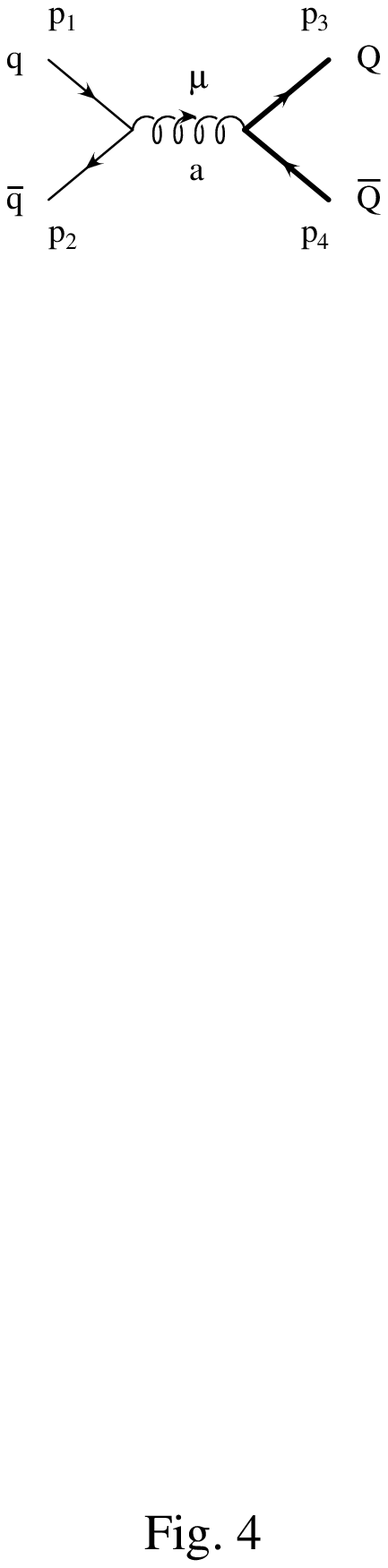}}
%\end{figure}

\newpage
%\begin{figure} 
\centering
\mbox{\epsffile{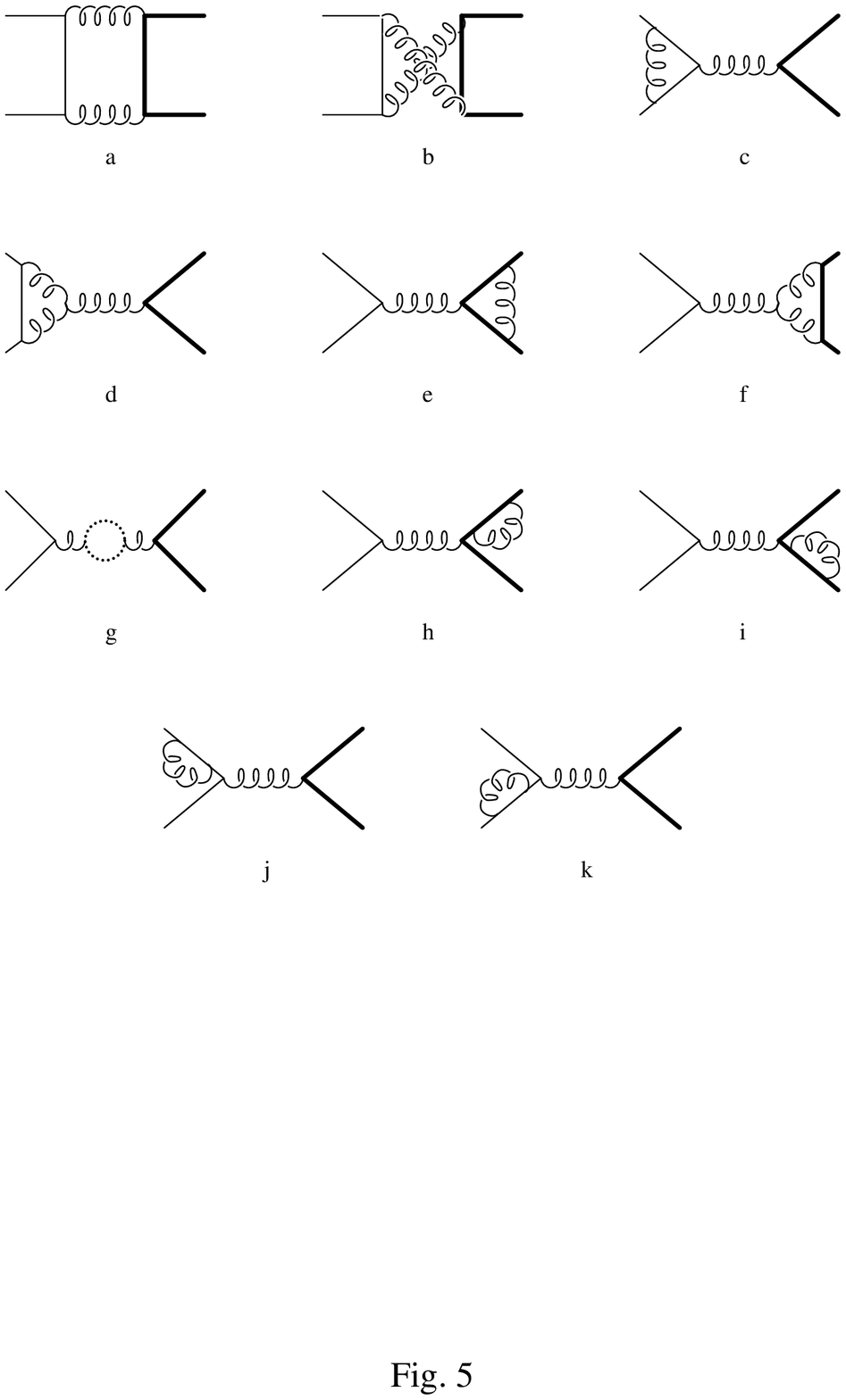}}
%\end{figure}

\end{document}